\documentclass{aa}
\usepackage{graphicx}
\usepackage{txfonts}
\usepackage{natbib}
\bibpunct{(}{)}{;}{a}{}{,}
\usepackage{lscape}
\begin{document} 
   \title{Candidate free-floating super-Jupiters in the young \\ $\sigma$~Orionis open cluster}
%
   \titlerunning{Candidate free-floating super-Jupiters in $\sigma$ Orionis}
   \author{G.~Bihain\inst{1}\fnmsep\inst{2}
      \and R.~Rebolo\inst{1}\fnmsep\inst{2}
      \and M.~R.~Zapatero~Osorio\inst{1}
      \and V.~J.~S.~B{\'e}jar\inst{1}
      \and I.~Vill{\'o}-P{\'e}rez\inst{3}
      \and A.~D\'{\i}az-S{\'a}nchez\inst{3}
      \and A.~P{\'e}rez-Garrido\inst{3}
      \and J.~A.~Caballero\inst{4}
      \and C.~A.~L.~Bailer-Jones\inst{5}
      \and D.~Barrado~y~Navascu{\'e}s\inst{6,7}
      \and J.~Eisl{\"o}ffel\inst{8}
      \and T.~Forveille\inst{9}
      \and B.~Goldman\inst{5}
      \and T.~Henning\inst{5}
      \and E.~L.~Mart\'{\i}n\inst{1,10}
      \and R.~Mundt\inst{5}
          }
    \authorrunning{Bihain}
    \institute{Instituto de Astrof\'{\i}sica de Canarias, c/ V\'{\i}a L\'actea, s/n, 38205  La Laguna, Tenerife, Islas Canarias, Spain\\
	     \email{gbihain@ll.iac.es}
	     \and
	     Consejo Superior de Investigaciones Cient\'{\i}ficas, Spain
	     \and
	     Universidad Polit{\'e}cnica de Cartagena. Campus Muralla del Mar, Cartagena, Murcia, E-30202, Spain
	     \and
             Dpto. de Astrof\'{\i}sica y Ciencias de la Atm\'osfera, Facultad de F\'{\i}sica, Universidad Complutense de Madrid, E-28040 Madrid, Spain
	     \and
	     Max-Planck-Institut f{\"u}r Astronomie, K{\"o}nigstuhl 17, D-69117 Heidelberg, Germany
	     \and
	     Laboratorio de Astrof\'{\i}sica Espacial y Exoplanetas, Centro de Astrobiologia (LAEFF-CAB, INTA-CSIC), European Space Astronomy centre (ESAC), PO Box 78, 28691 Villanueva de la Ca{\~n}ada, Madrid, Spain
	     \and
             Spanish Virtual Observatory thematic network, Spain
	     \and
	     Th{\"u}ringer Landessternwarte Tautenburg, Sternwarte 5, D-07778 Tautenburg, Germany
	     \and
	     Laboratoire d'Astrophysique de Grenoble, Observatoire de Grenoble, Universit{\'e} Joseph Fourier, CNRS, UMR 571 Grenoble, France
	     \and
	     University of Central Florida. Department of Physics, P.O. Box 162385, Orlando, FL 32816-2385, USA
             }
   
   \date{Received 26 March 2009 / Accepted 31 July 2009} 

   \abstract
    {Free-floating substellar candidates with estimated theoretical masses
of as low as $\sim$5 Jupiter masses have been found in the $\sim$3~Myr old
\object{$\sigma$~Orionis} open cluster. As the overlap with the planetary mass
domain increases, the question of how these objects form becomes
important. The determination of their number density and whether a mass
cut-off limit exists is crucial to understanding their formation.}  
    {We propose to search for objects of yet lower masses in the cluster
and determine the shape of the mass function at low mass.}
    {Using new- and (re-analysed) published $IZJHK_{\rm s}[3.6]-[8.0]$-band data of an area of
840~arcmin$^{2}$, we performed a search for LT-type cluster member
candidates in the magnitude range $J=19.5$--21.5~mag, based on their expected
magnitudes and colours.}
    {Besides recovering the T~type object \object{S\,Ori~70} and two other known
objects, we find three new cluster member candidates, S\,Ori~72--74, with
$J\approx21$~mag and within 12~arcmin of the cluster centre. They have theoretical
masses of 4$_{-2}^{+3}$~$M_{\rm Jup}$ and are among the least massive
free-floating objects detected by direct imaging outside the Solar System. The
photometry in archival {{\em Spitzer}} [3.6]--[5.8]-band images infers
that \object{S\,Ori~72} is an L/T transition candidate and \object{S\,Ori~73} a
T-type candidate, following the expected cluster sequence in the mid-infrared.
Finally, the L-type candidate \object{S\,Ori~74} with lower quality
photometry is located at 11.8~arcsec ($\sim$4250~AU) of a stellar member of
$\sigma$~Orionis and could be a companion. After contaminant correction in the
area complete to $J=21.1$~mag, we estimate that there remain between zero and two
cluster members in the mass interval 6--4~$M_{\rm Jup}$.} 
{We present S\,Ori~73, a new candidate T~type and candidate $\sigma$~Orionis
member of a few Jupiter masses. Our result suggests a possible turnover in the
substellar mass spectrum below $\sim$6 Jupiter masses, which could be
investigated further by wider and deeper photometric surveys.}


\maketitle

\section{Introduction}\label{intro}

Free-floating objects with masses of several to a few times the mass of Jupiter
appear to populate young open clusters
\citep[see][]{lucas2000,zapateroosorio2000}. They could form in a similar way to
stars, by gravitational fragmentation above an opacity mass limit
\citep{hoyle1953,larson1973,low1976,rees1976,silk1977} or by turbulent
fragmentation \citep{padoan2002,padoan2004,padoan2007} of collapsing molecular
clouds, or as stellar embryos that are fragmented, photo-eroded, or ejected before
they can accrete sufficient mass to become stars (see \citealt{whitworth2005} and
references therein). They could also form by gravitational instability in
circumstellar disks \citep{boss1997,whitworth2006}, then have their orbits
disrupted and be ejected \citep{stamatellos2009,veras2009}. A better knowledge of
the cluster mass function (MF; number of objects per unit mass) at these low
masses will help us to determine the main formation process for these objects.
Indeed, numerical simulations of opacity-limited fragmentation show a cutoff in
the mass function at $\sim$4 Jupiter masses \citep{bate2005,bate2005b,bate2009},
whereas numerical simulations of turbulent fragmentation show an approximately
log-normal, shallower drop at substellar masses \citep{padoan2004}. A detailed
comparison with planets (e.g., spectral emission and chemical composition) will
also provide complementary information about their origin and evolution
\citep{fortney2008}.

The \object{$\sigma$~Orionis} open cluster in the \object{Ori OB 1b} association,
together with other star-forming regions in the \object{Orion} and
Scorpius-Centaurus complexes, is well-suited to the search for free-floating
planetary-mass objects. It is young \citep[$3\pm2$~Myr;][]{zapateroosorio2002a},
relatively nearby ($360^{+70}_{-60}$~pc, \citealt{brown1994}; $444\pm20$~pc,
\citealt{sherry2008}), affected by very low extinction \citep[$A_{\rm V}<1$~mag;][]{sherry2008}, 
and of solar metallicity \citep{gonzalezhernandez2008}.
A revision of published, basic parameters of the cluster was provided by
\citet{caballero2007bri}. \citet{caballero2007} found a smoothly continuous MF
down to $\sim$6 Jupiter masses ($M_{\rm Jup}$) and that the brown dwarfs appear to
harbour disks with a frequency similar to that of low-mass stars. This suggests
that low-mass stars and substellar objects share the same formation mechanism.
Also, \object{S\,Ori~70}, of spectral type T6, has been proposed to be a cluster
member with an estimated mass of 2--7~$M_{\rm Jup}$
\citep{zapateroosorio2002b,zapateroosorio2008,martin2003a,burgasser2004,scholz2008,luhman2008}.

In this paper, we present new $IZJHK_{\rm s}$-band photometry and a re-analysis
of previous data of the \object{$\sigma$~Orionis} cluster, allowing us to search
for faint candidates in an area of $\sim$790~arcmin$^2$, to the completeness
magnitude $J\approx21.1$~mag. Our search area overlaps with those of
\citet{caballero2007} and \citet{lodieu2009somf}, and its $J$-band completeness
magnitude is about 1.5 and 2~mag fainter, respectively. We report the detection
of three new cluster member candidates with theoretical masses of
$\sim$4~$M_{\rm Jup}$.

\section{Observations and data reduction}

We discuss the new data obtained for this study and the data from
\citet{caballero2007} and \citet{zapateroosorio2008} that were reduced or analysed
again in an attempt to increase the sensitivity to faint sources.

\subsection{Optical data}\label{optd}

The $I$-band imaging data presented in \citet{caballero2007} were obtained with
the Wide Field Camera (WFC) mounted at the Isaac Newton Telescope (INT). The WFC
contains four CCD of 2\,k~$\times$~4\,k pixels and 0.33~arcsec/pixel.
Figure~\ref{map} shows the area of the corresponding four images, limited to their
overlap with the near-infrared data (solid and dashed lines). A new automatic
search for sources was performed with the {\tt IRAF} routine {\tt FINDSTAR}
(E.~Almoznino), which led to a substantial increase in the number of sources at
faint magnitudes with respect to those considered by \citet{caballero2007}. {\tt
FINDSTAR} is especially useful for detecting sources in combined dithered images
(or in images with background gradients), where the standard deviation varies from
centre to border. We then carried out the aperture and point-spread-function (PSF)
photometry using routines within  the {\tt DAOPHOT} package. Objects missed by the
automatic search routine but easily detected by eye in the PSF-subtracted images
(e.g., sources partially hidden in the wings of bright stars) were added to the
list of targets.  Finally, for each of the four CCD images, the new photometry was
calibrated using $\sim$1850~objects in common with the \citet{caballero2007}
photometry in the Cousins system. We found average completeness and limiting
magnitudes of $I_{\rm cmp}=23.0$~mag and $I_{\rm lim}=23.9$~mag, respectively.

\begin{figure}[ht!] \resizebox{\hsize}{!}{\includegraphics{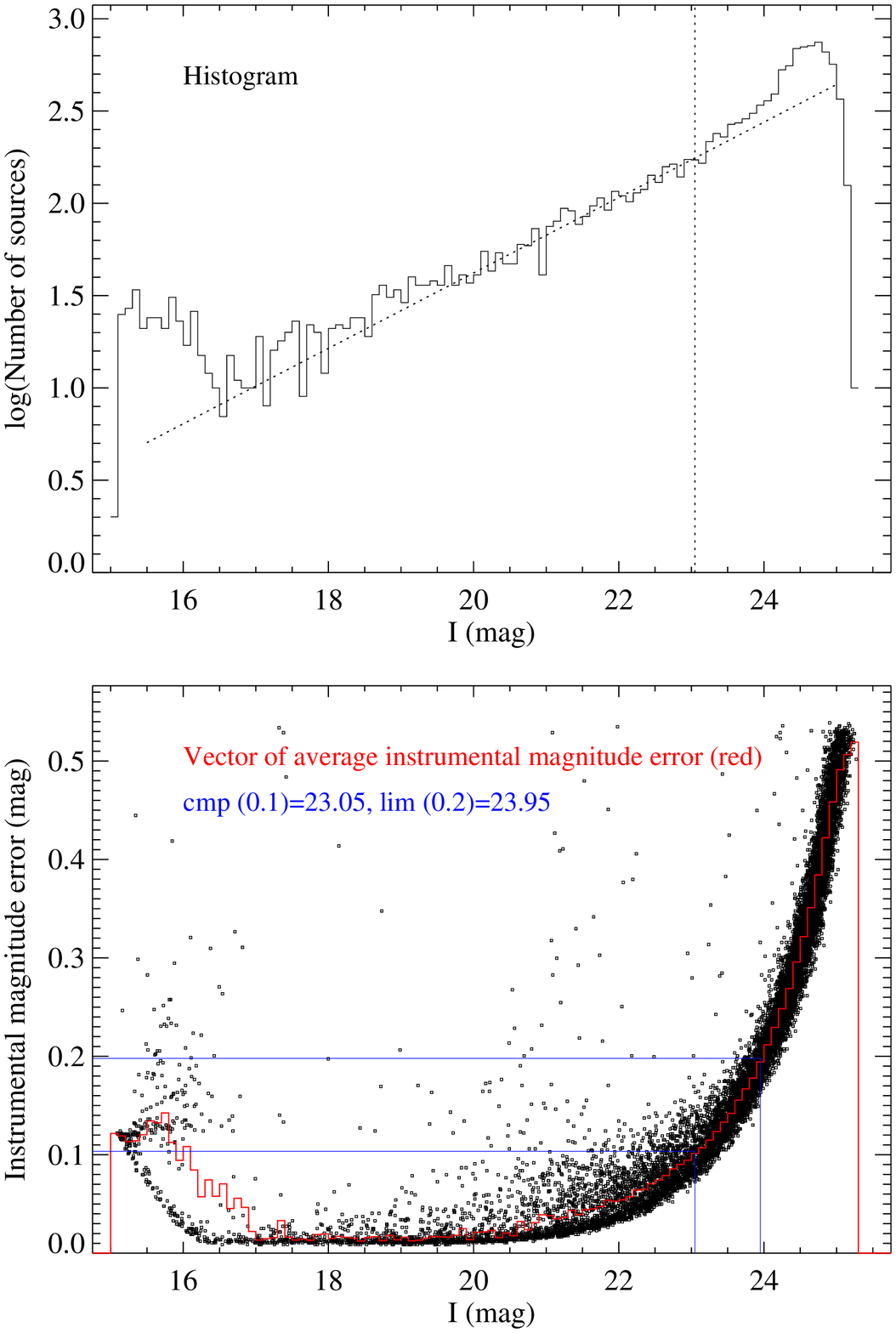}}
\caption{{\it Upper panel}. Histogram in logarithmic scale of the $I$-band
sources as a function of the calibrated magnitude (see text for details about
the dotted lines). {\it Lower panel}. Instrumental magnitude error versus the
calibrated magnitude of these sources (dots). The average error per magnitude
bin is overploted as a red solid line. The magnitudes of the bins just below
errors of 0.1~mag (blue lower line) and 0.2~mag (blue upper line) were defined
here as the completeness and limiting magnitudes, respectively.} 
\label{histvavg} \end{figure}

To determine these completeness and limiting magnitudes, we compiled the
distribution of the instrumental magnitude error versus the calibrated magnitude
for each image. In the bottom panel of Fig.~\ref{histvavg}, we present this with
the source catalogue of one of the WFC CCD images.  The completeness and
limiting magnitudes were defined to be the faintest magnitude bins where the
average errors are $\le$0.10 and 0.20~mag, respectively. These errors correspond
to signal-to-noise ratios of $S/N=10$ and $S/N=5$, respectively \citep[see e.g.,
][]{newberry1991}. The average error per magnitude bin is overploted as a red
solid line. The upper panel shows the histogram of all sources on a logarithmic
scale as a function of the calibrated magnitude. The inclined dotted line
represents a power law fit to the histogram in the range $[mag_{\rm cmp}-2.5,
mag_{\rm cmp}]$, where $mag_{\rm cmp}$ is the completeness magnitude (vertical
dotted line). At $mag>mag_{\rm cmp}$, the histogram's deviation from the fit is
most probably caused by sources affected by random upward or downward
fluctuations of the background, the upward ones being preferentially detected
above the detection threshold (Malmquist bias; see also \citealt{beichman2003}).
Comparing the counts of the histogram of sources having errors smaller than
0.10~mag with the counts of the linear extrapolation of its power law fit at the
completeness magnitude, we estimated a level of completeness of $\ga$90\,\%.
This method was also applied to the other data sets.

\begin{table}
\protect\caption[]{Coordinates and depth of the LRIS images$^{a}$.}
\label{IZlris}
\centering          
     \begin{tabular}{c c c c}
     	 \hline\hline
	 ID &
	 $\alpha$\,\,\,\,\,\,\,\,\,\,\,\,\,\,\,\,\,\,\,\,\,\,\,\,\,\,\,\,$\delta$ &
	 Filter &
	 m$_{\rm cmp}$, \,\,m$_{\rm lim}$
         \\
	 & (J2000)\,\,\,\,\,\,\,\,\,\,\,\,\,(J2000) & & (mag)
	 \\
	 \hline
	 1998i1  & 05 38 16.6 \,\,\,$-$02 37 53 & $I$ & 23.6, \,\,24.5 \\
	 1998i2  & 05 38 21.0 \,\,\,$-$02 47 59 & $I$ & 23.5, \,\,24.4 \\
	 1998i3  & 05 38 56.4 \,\,\,$-$02 38 57 & $I$ & 23.5, \,\,24.1 \\
	 1998i4  & 05 39 21.2 \,\,\,$-$02 37 54 & $I$ & 23.5, \,\,24.4 \\
	 1998i5  & 05 39 56.4 \,\,\,$-$02 34 00 & $I$ & 23.6, \,\,24.4 \\
	 1998z1  & 05 38 16.6 \,\,\,$-$02 37 53 & $Z$ & 22.1, \,\,22.9 \\
	 1998z2  & 05 38 21.0 \,\,\,$-$02 47 59 & $Z$ & 22.4, \,\,23.3 \\
	 1998z3  & 05 38 56.4 \,\,\,$-$02 38 57 & $Z$ & 22.4, \,\,23.3 \\
	 1998z4  & 05 39 21.2 \,\,\,$-$02 37 54 & $Z$ & 22.5, \,\,23.4 \\
	 1998z5  & 05 39 56.4 \,\,\,$-$02 34 00 & $Z$ & 22.4, \,\,23.3 \\
	 2000i1  & 05 38 17.4 \,\,\,$-$02 37 48 & $I$ & 23.8, \,\,24.6 \\
	 2000i2  & 05 38 21.4 \,\,\,$-$02 48 03 & $I$ & 23.5, \,\,24.5 \\
	 2000i3  & 05 38 55.2 \,\,\,$-$02 38 52 & $I$ & 23.3, \,\,24.3 \\
	 2000i4  & 05 39 20.3 \,\,\,$-$02 37 58 & $I$ & 23.7, \,\,24.5 \\
	 2000i5  & 05 39 57.2 \,\,\,$-$02 34 04 & $I$ & 23.6, \,\,24.7 \\
	\hline
      \end{tabular}
\begin{flushleft}
$^{a}$ The pixel scale is 0.210~arcsec~${\rm pixel}^{-1}$ and the field
of view in the $\alpha$--$\delta$ frame is 5.8$\times$7.2$\sim$arcmin$^{2}$. For the
images 1998[i,z]3, 2000i[1, 2,4,5], and 2000i3, this field of view is rotated
$\sim$16, 90, and 105~deg counterclockwise, respectively.
\end{flushleft}
\normalsize
\end{table}

\begin{table*}
\protect\caption[]{New near-infrared observations$^a$.}
\label{obslog}
\centering          
     \begin{tabular}{c c c c c c c c c c}
     	 \hline\hline
         ID &
	 $\alpha$\,\,\,\,\,\,\,\,\,\,\,\,\,\,\,\,\,\,\,\,\,\,\,\,\,\,\,\,$\delta$ &
	 Instr. &
	 Filter &
	 Area &
	 Date(s) &
	 t$_{\rm exp}$ &
	 m$_{\rm cmp}$ &
	 m$_{\rm lim}$
         \\
	 & (J2000)\,\,\,\,\,\,\,\,\,\,\,\,\,(J2000) &&& (arcmin$^2$)&& (min) & (mag) & (mag)
	 \\
	 \hline
        1-k & 05 39 39.9 \,\,\,$-$02 50 26  & Omega2000 & $K_{\rm s}$ &197 (279)& 2005 Jan 31, Feb 1, Oct 26& 326   & 20.5 & 21.2 \\

        2-j & 05 39 35.1 \,\,\,$-$02 34 44  & Omega2000 & $J$	      &214 (259)& 2006 Oct 07		    &  63   & 20.9 & 21.9 \\
        2-h & 05 39 19.3 \,\,\,$-$02 33 36  & Omega2000 & $H$	      &181 (295)& 2005 Oct 19		    & 186   & 21.2 & 21.9 \\
        2-k & 05 39 21.2 \,\,\,$-$02 35 14  & Omega2000 & $K_{\rm s}$ &194 (282)& 2005 Oct 24--25	    & 139.5 & 20.1 & 21.0 \\

        3-h & 05 38 40.1 \,\,\,$-$02 48 09  & SofI	& $H$	      & 21 (27)  & 2006 Dec 27  	    & 147   & 20.3 & 20.9 \\
        4-h & 05 38 40.0 \,\,\,$-$02 52 26  & SofI      & $H$	      & 21 (27)  & 2006 Dec 24		    &  98   & 20.3 & 21.1 \\
        5-h & 05 38 39.7 \,\,\,$-$02 57 09  & SofI      & $H$	      & 21 (27)  & 2006 Dec 27		    &  75   & 20.1 & 20.5 \\
        6-h & 05 38 20.0 \,\,\,$-$02 48 30  & SofI      & $H$	      & 21 (27)  & 2006 Dec 24		    & 201   & 20.1 & 21.1 \\
        7-h & 05 40 05.1 \,\,\,$-$02 30 40  & SofI      & $H$	      & 21 (27)  & 2006 Dec 27		    & 132   & 20.3 & 21.1 \\
        8-h & 05 40 04.9 \,\,\,$-$02 35 57  & SofI      & $H$	      & 21 (27)  & 2006 Dec 26		    &  98   & 19.5 & 20.1 \\
        9-h & 05 40 03.6 \,\,\,$-$02 42 10  & SofI      & $H$	      & 21 (27)  & 2006 Dec 25		    &  56   & 19.9 & 20.7 \\
       10-h & 05 39 09.7 \,\,\,$-$02 47 15  & SofI      & $H$	      & 21 (27)  & 2006 Dec 25		    & 168   & 20.3 & 21.3 \\
       11-h & 05 39 27.7 \,\,\,$-$02 47 15  & SofI      & $H$	      & 21 (27)  & 2006 Dec 25		    &  95   & 20.1 & 21.1 \\
       12-h & 05 39 45.7 \,\,\,$-$02 47 15  & SofI      & $H$	      & 21 (27)  & 2006 Dec 26		    & 120   & 20.3 & 21.3 \\
       13-h & 05 40 03.1 \,\,\,$-$02 47 04  & SofI      & $H$	      & 16 (33)  & 2006 Dec 26		    & 140   & 20.3 & 21.1 \\

       14-h & 05 39 09.5 \,\,\,$-$02 53 56  & LIRIS	& $H$	      & 16 (20)	 & 2006 Dec 29		    &  36   & 19.5 & 20.1 \\
       15-h & 05 39 45.4 \,\,\,$-$02 53 57  & LIRIS	& $H$	      & 16 (20)  & 2007 Dec 14		    &  45   & 20.3 & 20.9 \\
       16-h & 05 40 03.6 \,\,\,$-$02 53 59  & LIRIS	& $H$	      & 16 (20)  & 2007 Dec 14		    &  45   & 20.3 & 20.9 \\
	\hline
      \end{tabular}
\begin{flushleft}
$^a$ Fields of view (and pixel scales) of the Omega2000, SofI, and LIRIS detectors are $15.4\times15.4$~arcmin$^2$ (0.45~arcsec~${\rm pixel}^{-1}$),
$4.9\times4.9$~arcmin$^2$ (0.288~arcsec~${\rm pixel}^{-1}$), and $4.2\times4.2$~arcmin$^2$ (0.25~arcsec~${\rm
pixel}^{-1}$), respectively.\\
\end{flushleft}
\normalsize
\end{table*}

We used broad $IZ$-band images from the Keck~II Low Resolution Imaging
Spectrograph (LRIS), associated with the discovery of S\,Ori~70
\citep[see][]{zapateroosorio2002b}, as well as unpublished $I$-band images
obtained with the same instrument on 2000 January 5, in similar atmospheric
conditions and for the same exposure time. Their coordinates and depths are
listed in Table~\ref{IZlris}. The photometry was performed as described above.
The $I$-band photometry was calibrated using typically 200 sources in common
with the WFC survey, whereas the $Z$-band photometry was calibrated using on
average 41 stellar sources from the Galactic Clusters Survey (GCS) component of
UKIDSS\footnote{UKIDSS uses the UKIRT Wide Field Camera (WFCAM;
\citealt{casali2007}) and a photometric system described in \citet{hewett2006}.
The pipeline processing and science archive are described in
\citet{hambly2008}.} \citep[][fifth data release]{lawrence2007} and of magnitude
error $\sigma_{Z~(\rm UKIDSS)}<0.1$~mag. The $Z$-band photometry has an average
(relative) calibration error of 0.04~mag. We caution that the UKIDSS Z-band
filter differs from that of the LRIS images. For the published data, we
estimated that the completeness and limiting magnitudes are $I_{\rm
cmp}=23.5$~mag and $I_{\rm lim}=24.3$~mag, respectively, and $Z_{\rm
cmp}=22.3$~mag and $Z_{\rm lim}=23.2$~mag, respectively. For the data from 2000,
these are $I_{\rm cmp}=23.5$~mag and $I_{\rm lim}=24.5$~mag.

We obtained $Z$-band imaging data using the INT/WFC instrument on the night of
2008 November 27. We took 21~images of 900~s exposure time each with central
coordinates $(\alpha, \delta)=(5~37~49.1, -2~44~43)$. During the observations,
thin cirrus were present and the average seeing was 1.2~arcsec. The images were
reduced using routines within the {\tt IRAF} environment, including bias and
zero image subtraction and flat-field correction. Observations were done using a
dithering pattern. Science images were combined to obtain flat-field images to
correct for fringing. Individual images were aligned and combined to obtain
final images. Aperture and PSF photometry was performed for one of the CCDs. Its
photometric calibration was done using about 400 stellar sources from GCS-UKIDSS
of $\sigma_{Z~(\rm UKIDSS)}<0.1$~mag, implying a relative calibration error
of 0.02~mag. We caution that the UKIDSS Z-band filter is different from that of
the WFC images. We estimated completeness and limiting magnitudes of 22.4 and
23.1~mag, respectively.

Astrometry was obtained for all the optical images with an accuracy
of $\sim$0.2--0.05~arcsec, using 2MASS as reference and an adaptation of the
{\tt IRAF MYASTROM} procedure (E. Puddu; see also \citealt{bihain2006}). A
representation of the individual fields is provided in Fig.~\ref{mapsi} (top
left and right panels).

\subsection{Near- and mid-infrared data}\label{nIRphot}

\begin{figure}[ht!]
\resizebox{\hsize}{!}{\includegraphics{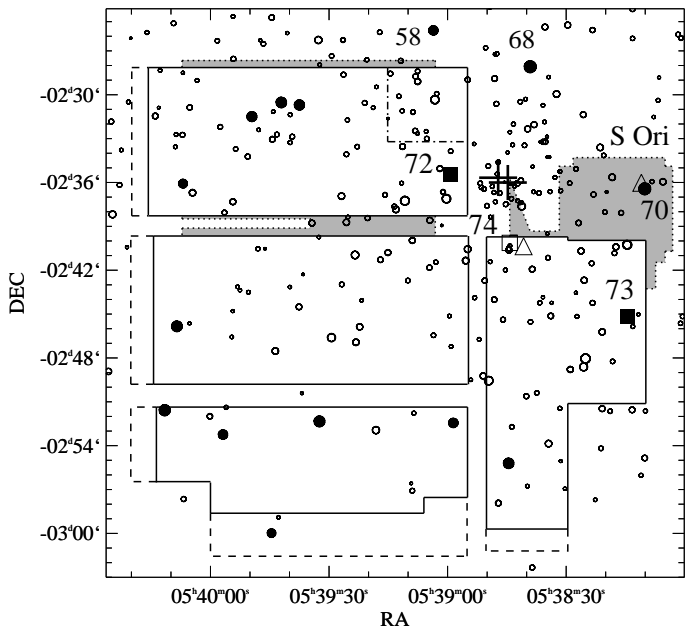}}
\caption{Main search area: WFC-ISAAC $IJ$-band data (dashed line)
together with follow-up $H$- or $K$-band data (solid line). Additional search
areas: WFC $I$-, LRIS $I$-, and Omega2000 $JHK_{\rm s}$-band data (shaded regions
delimited by dotted lines). Individual fields are not represented for clarity
(see Fig.~\ref{mapsi}). These areas have a completeness $J\ge21.1$~mag, except
the upper corner in the main area (indicated by the dash-dot line) and the
shaded regions on the left side. \object{S\,Ori~72} and \object{S\,Ori~73} are
represented by filled squares, \object{S\,Ori~74} by an open square, and
\object{S\,Ori~J053840.8$-$024022} and \object{S\,Ori~J053811.0$-$023601} by
open triangles (see finding charts of Figs.~\ref{fcharts72}, \ref{fcharts73},
\ref{fcharts74glx1}, and \ref{fchartsglx2}). Circular symbols are cluster
members and candidates
\citep{zapateroosorio2000,caballero2007,caballero2008mayrit}; their size
increases with {\em fainter} $J$-band magnitude, to highlight the location of
the least massive objects. Filled circles are planetary-mass candidates with
$M<0.013M_{\odot}$, including \object{S\,Ori~70} \citep{zapateroosorio2002b}.
Crosses are \object{$\sigma$~Ori~AB} and E, at the centre of the cluster.} 
\label{map} \end{figure}

The $J$-band imaging data from \citet{caballero2007} were obtained with the
Infrared Spectrometer And Array Camera (ISAAC), mounted at the Very Large
Telescope (VLT) and containing a Rockwell Hawaii detector of 1\,k~$\times$~1\,k
pixels and 0.148~arcsec/pixel. We re-reduced these data to obtain a clean sky
subtraction, remove bad pixel values, and identify more reliably charge
persistencies of bright sources in the detector. The raw images were dark
subtracted, superflat divided, sky subtracted (with the routine {\tt
LIRISDR.LIMAGE.LRUNSKY} from J.~A. Acosta-Pulido, which includes object masking
and vertical gradient correction), their elements flagged for bad pixel (at
extreme values for thresholding) using bad pixel masks from superflats, related
by pixel shifts (computed from clearly defined sources in common), and combined
all at once in strips along right ascension or declination (16 strips in total).
The photometry of each strip was performed similarly as for the WFC data. We
ensured that all the sources remaining in the PSF-subtracted images were
recovered, as we did for the Omega2000 $J$-band images (see below). The
photometry was calibrated using an average number of 12 point sources from the
2MASS catalogue \citep{skrutskie2006} of quality flags AAA or AAB. The average
calibration error is 0.03~mag. The average completeness and limiting magnitudes
are $J_{\rm cmp}=21.6$~mag and $J_{\rm lim}=22.4$~mag, respectively, in an area
of $\sim$660~arcmin$^{2}$. This area excludes $\sim$20~arcmin$^{2}$ within the
region delimited by the dash dot line in Fig.~\ref{map}, where $J_{\rm
cmp}=20.6$~mag and $J_{\rm lim}=21.5$~mag, and $\sim$100~arcmin$^{2}$
corresponding to the $\sim$0.4~mag shallower borders of the strips.

We obtained additional near-infrared imaging data, using Omega2000 at the 3.5-m
Telescope (Calar Alto, Spain), Son of Isaac (SofI) at the New Technology
Telescope (La Silla, Chile), and the Long-slit Intermediate Resolution Infrared
Spectrograph (LIRIS) at the William Herschel Telescope (Roque de los Muchachos
Observatory, Spain). Table~\ref{obslog} indicates field identification,
coordinates, instrument, filter, area, observing night date(s), total exposure
time, and completeness- and limiting magnitudes. All the data were reduced
within the {\tt IRAF} environment, including (super)flat division, sky
subtraction, alignment with several reference stars, and combination without
trimming. SofI raw images were first row cross-talk corrected with the routine
{\tt crosstalk} (Leo Vanzi, ESO SofI tools). Omega2000 and most of the SofI
images were dark subtracted before flat division. LIRIS raw images were first
pixel-mapping- and row-cross-talk corrected, and then processed with the routine
{\tt LIRISDR.LIMAGE.LDEDITHER}, including sky subtraction, as applied to
ISAAC $J$-band, and distortion correction. Bad pixel masks were used for LIRIS
and SofI, whereas for Omega2000 extreme values relative to the average at each
pixel were rejected during combination. Because the combined images are
untrimmed, they have a deep central region surrounded by a shallower region,
whose proportions depend on the observing dithers. In Table~\ref{obslog}, we
list the deep area and the total area (in parenthesis). The photometry was
obtained as described above. For the Omega2000 images, the photometric
calibration was obtained using $\sim$150 2MASS point sources of quality flags
AAA or AAB (average calibration error of 0.03~mag). For each of the SofI and
LIRIS images, about 10 of these calibrators were used (average calibration error
of 0.04~mag). Completeness and limiting magnitudes were derived from the sources
in the deeper central regions.

Other near-infrared data, already published in \citet{caballero2007} and
\citet{zapateroosorio2008}, were used in the search. The Omega2000 data from the
latter study were re-reduced to obtain untrimmed images. Completeness and
limiting magnitudes of our new photometry (obtained as described above) are
listed in Table~\ref{HKother}, except for the $H$-band $\sim$1100~arcmin$^2$
Omega2000 survey from 2003 (see \citealt{caballero2007}), which is shallower. We
note that the field 1-k is a combination of new data obtained on 2005 October 26
with published data obtained earlier the same year, on January 31 and February
1. The $J$- and $HK$-band data from Tables~\ref{obslog} and \ref{HKother}
correspond to overlapping areas of $\sim$240~arcmin$^2$ and $\sim$690~arcmin$^2$
with the WFC~+~ISAAC survey, respectively (see Fig.~\ref{map} for the $HK$-band
data).

\begin{table}
\protect\caption[]{Re-estimated depth of individual fields from \citet{caballero2007} and
\citet{zapateroosorio2008}.}
\label{HKother}
\centering          
     \begin{tabular}{c c c c c}
     	 \hline\hline
	 ID &
	 $\alpha$\,\,\,\,\,\,\,\,\,\,\,\,\,\,\,\,\,\,\,\,\,\,\,\,\,\,\,\,$\delta$ &
	 Instr. &
	 Filter &
	 m$_{\rm cmp}$, \,\,m$_{\rm lim}$
         \\
	 & (J2000)\,\,\,\,\,\,\,\,\,\,\,\,\,(J2000) &&& (mag)
	 \\
	 \hline
 1-h & 05 39 37.6 \,\,\,$-$02 49 53 & Omega2000  & $H$   	    & 20.0, \,\,20.9  \\

17-h & 05 38 38.7 \,\,\,$-$02 49 01 & Omega2000  & $H$  	    & 19.1, \,\,19.6  \\
17-k & 05 38 37.6 \,\,\,$-$02 49 51 & Omega2000  & $K_{\rm s}$      & 19.6, \,\,20.3  \\

18-j$^{a}$ & 05 38 12.4 \,\,\,$-$02 35 18 & Omega2000 & $J$	    & 21.1, \,\,21.9  \\
18-h$^{a}$ & 05 38 11.9 \,\,\,$-$02 34 59 & Omega2000 & $H$	    & 20.6, \,\,21.4  \\
18-k$^{a}$ & 05 38 13.3 \,\,\,$-$02 35 37 & Omega2000 & $K_{\rm s}$ & 20.0, \,\,21.0  \\

19-h & 05 38 44.6 \,\,\,$-$02 44 35 & CFHTIR  & $H$	    & 21.1, \,\,22.1  \\
19-k & 05 38 44.6 \,\,\,$-$02 44 35 & CFHTIR  & $K'$	    & 20.9, \,\,21.9  \\
20-h & 05 38 18.3 \,\,\,$-$02 44 31 & CFHTIR  & $H$	    & 20.5, \,\,21.3  \\
20-k & 05 38 18.3 \,\,\,$-$02 44 31 & CFHTIR  & $K'$	    & 20.5, \,\,21.3  \\
21-h & 05 39 24.2 \,\,\,$-$02 29 36 & CFHTIR  & $H$	    & 20.9, \,\,21.5  \\
21-k & 05 39 25.6 \,\,\,$-$02 29 37 & CFHTIR  & $K'$	    & 20.5, \,\,21.3  \\
22-h & 05 40 06.5 \,\,\,$-$02 32 29 & CFHTIR  & $H$	    & 20.7, \,\,21.9  \\
22-k & 05 40 06.6 \,\,\,$-$02 32 29 & CFHTIR  & $K'$	    & 20.5, \,\,21.5  \\
	\hline
      \end{tabular}
\begin{flushleft}
$^{a}$ Re-reduced data from \citet{zapateroosorio2008}; deep central areas (and total areas) in $J$, $H$, and
$K_{\rm s}$ are 215 (257), 208 (265), and 197 (278)~arcmin$^2$. For the Omega2000 and CFHTIR data from
\citet{caballero2007}, the deep central areas (and total areas) are 216 (236) and 6 (22)~arcmin$^2$,
respectively.
\end{flushleft}
\normalsize
\end{table}

Astrometry was obtained for all the near-infrared images similarly as for the
optical images, with an accuracy of $\sim$0.2--0.05~arcsec. A representation of
the individual fields is provided in Fig.~\ref{mapsi} (top left and bottom
panels).

We also used archival post-basic calibrated data (PBCD) from the {\em Spitzer
Space Telescope} Infrared Array Camera (IRAC). For our new candidates (see
Sect.~\ref{resd}), we have obtained the Spitzer photometry following the
procedure described in \citet{zapateroosorio2007} and using the data published
by \citet{hernandez2007} and \citet{scholz2008}. A comparison of these two data
sets is provided in \citet[][see e.g. Fig.~1 therein for a map of the IRAC
surveys]{luhman2008}. We averaged our measurements in overlapping deep images
and adopted their standard deviation as a representative error bar. We compared
the [3.6]- and [4.5]-band measurements of \citet{zapateroosorio2007} with those
of \citet{luhman2008} for the six objects in common\footnote{\object{S Ori 54}
was not included in the comparison because in \citet{luhman2008} the object was
probably misidentified with the brighter source \object{[SE2004] 26, at about
5~arcsec}.} and found small differences $[3.6]_{\rm ZO - L}=0.02\pm0.12$~mag and
$[4.5]_{\rm ZO - L}=0.12\pm0.07$~mag, implying good agreement between the two
sets of measurements.

\section{The search for $\sigma$~Orionis LT-type objects} \label{tsearch}

Field dwarfs with spectral types T0--8 (effective temperature 1400--700 K) have
typical colours of $I-J>4.5$, $J-H$~$<$~1.5, and $J-K_{\rm s}<2$~mag
\citep{tinney2003,zhang2009}; the early types have redder $J-H$ and
$J-K_{\rm s}$ colours and higher effective temperatures than the later types.
By extrapolating the $\sigma$~Orionis cluster sequence using the field dwarf
sequence, cluster members with a T spectral type appear to be at $J\sim20$~mag
(see Sect.~\ref{resd} and Fig.~\ref{3p6}). About the same apparent magnitude is
found using the synthetic atmosphere $J$-band prediction of the 3-Myr COND model
isochrone from \citet{chabrier2000a}. However, when predicted bolometric
luminosities and effective temperatures are transformed into the observable
using relations for field dwarfs (see also Sect.~\ref{resd} and Fig.~\ref{3p6}),
a $J$-band value of $\sim$21~mag is found.

T-type objects of this magnitude will still be detected within the completeness
of the ISAAC data, whereas they will be relatively faint or undetected in the
less deep $IHK$-band images. For example, faint T~type objects with $J=21.5$,
$I-J>4.5$, $J-H<1$, and $J-K_{\rm s}<1$~mag will be undetected in all the
optical images and only possibly detected in the near-infrared images of
($H$- or $K_{\rm s}$-band) limiting magnitudes fainter than 20.5~mag
($\sim$470~arcmin$^2$). Therefore, we opted for a search relying on the ISAAC
$J$-band photometry, i.e., the deepest near-infrared photometry over the largest
area, and with an automatic selection in terms of magnitudes and colours that is
not too restrictive, to allow us to recover visually any potential cluster
member candidate, including L-type objects.

\begin{figure}[ht!] \resizebox{\hsize}{!}{\includegraphics{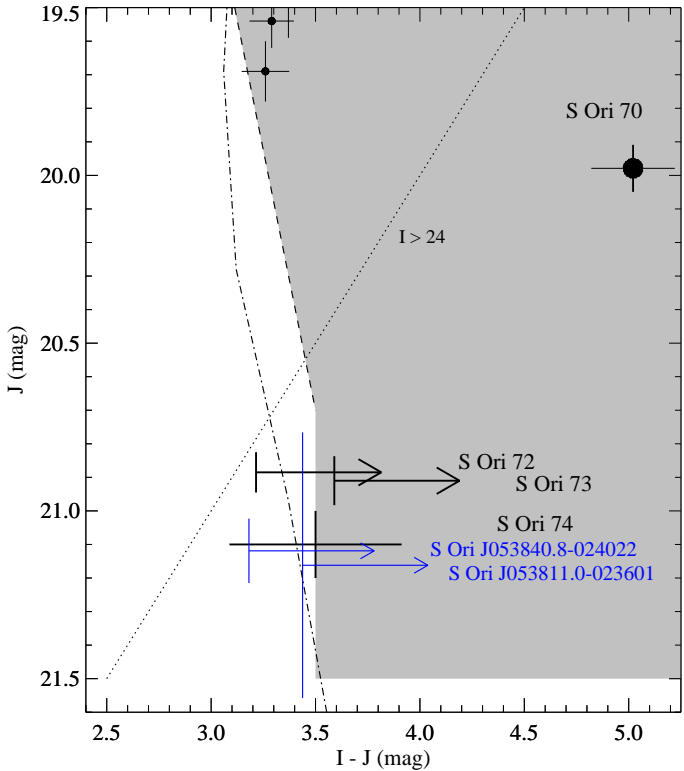}}
\caption{$J$ versus $I-J$ colour-magnitude diagram. The shaded search region is
where we expected cluster member candidates. The dashed line represents the
extrapolated selection boundary from \citet{caballero2007}. The dotted line
represents the $I>24$~mag selection sub-criterion. The dash-dot line represents
the 3~Myr COND model isochrone, where $I$ and $J$ are in the Cousins and CIT
photometric systems, respectively. The two small filled circles are the faintest
objects from \citet{caballero2007} and the large filled circle is S\,Ori~70. The
four (5~$\sigma$)~flux upper limits and one (1~$\sigma$)~error cross represent
our new candidates. Probable galaxies are highlighted in blue.}  \label{IJ}
\end{figure}

First, we correlated the $\alpha-\delta$ coordinates of the $IJHK_{\rm
s}$-band sources using the {\tt IDL srcor} procedure \citep[{\tt IDL} Astronomy
User's Library,][]{landsman1993}; for each $J$-band source, we searched for the
nearest counterpart within 2 arcsec in the $H$, $K_{\rm s}$, and $I$ bands. The
correlations with the WFC- and LRIS $I$-band catalogues were performed
separately. We then selected $19.5<J<21.5$~mag sources with no automatic
$I$-band detection, or either $I>24$~mag\,$\approx I_{\rm lim}$ (for unreliable
or spurious detections) or $I-J>3.5$~mag. As shown in Sect.~\ref{resd}, the
$I-J$ colour is essential for distinguishing LT-type objects from galaxies. The
$I-J>3.5$~mag sub-criterion intersects at $J=20.7$~mag with a linear
extrapolation of the selection criterion applied by \citet[][see therein
Fig.~2]{caballero2007} for their sources with $I<I_{\rm cmp}=23$~mag. For
continuity between the searches, we also selected sources redder than their
$I-J$ selection boundary and bluer than 3.5~mag. In the $J$ versus $I-J$
colour-magnitude diagram of Fig.~\ref{IJ}, the shaded region represents the
entire domain where we expected cluster member candidates, the dashed line
represents the extrapolated selection boundary from \citet{caballero2007}, and
the dotted line the $I>24$~mag sub-criterion. Finally, since the ``2-j''
Omega2000 $J$-band image (Table~\ref{obslog}) overlaps with the northern ISAAC
scans over $\sim$210~arcmin$^{2}$, we performed the selection process again for
the sources with $J_{\rm Omega2000}-J_{\rm ISAAC}>0.2$~mag and those without
ISAAC counterparts.

About 800 sources were chosen by our selection criteria from our $IJ$-band
catalogues. We checked each source visually in all of the optical and infrared
images, using the SAOImage DS9 display programme \citep{joye2003} and commands
in the X Public Access (XPA) messaging
system\footnote{http://hea-www.harvard.edu/saord/xpa/}. The simultaneous
visualisation in all available bandpasses and at all observing epochs allowed us
to verify whether a source is real (or of low proper motion) and unresolved.
Most sources were not detected automatically in the optical, because they are
faint or very close\footnote{De-blended in the images subtracted by the PSF
fitted sources ({\tt NOAO.DIGIPHOT.DAOPHOT.ALLSTAR}) or subtracted by a
smoothing obtained with a moving average box much smaller than the image size
({\tt NOAO.IMRED.CCDRED.MKSKYCOR}).} to brighter ones, and their clearly bluer
$I-J$ colours imply that they should be stars or unresolved galaxies. Many are
spurious detections of spikes or glares in the $J$-band. Others represent charge
persistencies of bright sources in the ISAAC detector (following precisely and
chronologically the offsets of individual pointings), resolved galaxies, sources
cut at image borders, or very blended sources, which are too close to bright
stars in the optical to be identified. The ``2-j'' Omega2000 $J$-band sources
with $J_{\rm Omega2000}-J_{\rm ISAAC}>0.2$~mag were typically galaxies, resolved
in the ISAAC images, whereas those without ISAAC counterparts were sources that
could not be detected in the shallower survey region (see Sect.~\ref{nIRphot}
and Fig.~\ref{map}) and the gaps between the strips.

In a similar way, we searched for candidates in the additional areas of
$\sim$15~arcmin$^{2}$ and $\sim$45~arcmin$^{2}$ represented by the shaded left
and right regions in Fig.~\ref{map}. These areas are common to the Omega2000
$JHK_{\rm s}$-, WFC $I$-, and LRIS $I$-band data. The $J$-band data of the left
and right regions (fields 2-j and 18-j) are complete to 20.9 and 21.1~mag,
respectively. They are therefore shallower by about 0.5~mag than the ISAAC data.

These searches allowed us to find four sources that are indeed undetectable by
eye in the deepest $I$-band images (see Sect.~\ref{resd}), and a half-dozen of
sources at $J>20.7$~mag that are barely detected beyond the $I$-band
limiting magnitudes. Most of the latter sources appear to be bluer than
$I-J\sim3.5$~mag. They have magnitude errors $\la$0.1~mag in the $JHK_{\rm
s}$-bands and red colours of $J-H\approx1$~mag or $J-K_{\rm s}\approx2$~mag.
Only one of them has a colour $I-J\ge3.5$~mag. It was selected as a candidate
(see Sect.~\ref{resd}), whereas the others were rejected because they are
probable galaxies or faint field M- or early L-type dwarfs. Some sources could
not be verified in the $I$-band images because of blending with extended stellar
spikes and glares. We estimated that areas of $\sim$10 and $\sim$5~arcmin$^{2}$
are lost in the main- and additional areas, respectively. Thus, the total search
area with $J$-band completeness $\ge$21.1~mag (ISAAC and Omega2000 18-j data)
amounts to $\sim$790~arcmin$^{2}$.

\section{Results and discussion}\label{resd}

\begin{table*}
\protect\caption[]{Coordinates and photometry of the new L- and T-type cluster member candidates$^a$.}
\label{LTcan}
\centering          
     {\footnotesize	
     \begin{tabular}{c c c c c c c c c c}
     	 \hline\hline
	 Name &
	 $\alpha$\,\,\,\,\,\,\,\,\,\,\,\,\,\,\,\,\,\,\,\,\,\,\,\,\,$\delta$ &
	 $I$ &
	 $Z$ &
	 $J$ &
	 $H$ &
	 $K_{\rm s}$ &
	 $[3.6]$ &
	 $[4.5]$ &
	 $[5.8]$
         \\
	 & (J2000)\,\,\,\,\,\,\,\,\,\,(J2000) & (mag) & (mag) & (mag) & (mag) & (mag) & (mag) & (mag) & (mag)
	 \\
	 \hline
         \object{S\,Ori~72} & 5 38 59.17 $-$2 35 26.0 & $>$24.1      & $>$23.3 & 20.89$\pm$0.06 & 19.57$\pm$0.03 & 18.72$\pm$0.05 & 17.65$\pm$0.11 & 17.51$\pm$0.29  & 17.28$\pm$0.38 \\
         \object{S\,Ori~73} & 5 38 14.49 $-$2 45 11.8 & $>$24.5 & 23.5$\pm$0.5 & 20.91$\pm$0.07 & 20.83$\pm$0.12 & 20.91$\pm$0.15 & 19.64$\pm$0.35 & 18.77$\pm$0.35  & $>$16.0        \\
         \object{S\,Ori~74} & 5 38 44.27 $-$2 40 07.9 & 24.6$\pm$0.4 & $>$23.3 & 21.1$\pm$0.1   & $>$18.54$^{b}$ & 19.38$\pm$0.10 & ...$^{c}$	   & ...$^{c}$       & $>$16.0        \\
	\hline
      \end{tabular}
\begin{flushleft}
$^a$ All have $[8.0]>15.1$~mag.
$^b$ Lower magnitude limit computed for the UKIDSS GCS specific field.
$^c$ Blended with a spike from the star \object{Mayrit~260182} (located at 11.8~arcsec).
\end{flushleft}
      }
\normalsize
\end{table*}

\begin{table*}
\protect\caption[]{Coordinates and photometry of probable galaxy candidates$^a$.}
\label{gcan}
     \begin{tabular}{c c c c c c c c c c}
     	 \hline\hline
	 Name &
	 $I$ &
	 $Z$ &
	 $J$ &
	 $H$ &
	 $K_{\rm s}$ &
	 $[3.6]$ &
	 $[4.5]$ &
	 $[5.8]$ &
	 $[8.0]$
	 \\
	 & (mag) & (mag) & (mag) & (mag) & (mag) & (mag) & (mag) & (mag) & (mag)
	 \\
	 \hline
         \object{S\,Ori~J053840.8$-$024022} & $>$24.3	  & ...     & 21.12$\pm$0.10 & 19.92$\pm$0.07 & 18.94$\pm$0.08 & 16.62$\pm$0.18 & 16.29$\pm$0.19 & $>$16.0	  & $>$15.1	   \\
         \object{S\,Ori~J053811.0$-$023601} & $>$24.6     & $>$22.9 & 21.16$\pm$0.40 & 20.66$\pm$0.11 & 19.61$\pm$0.09 & 17.16$\pm$0.02 & 16.53$\pm$0.04 & 15.56$\pm$0.06 & 14.69$\pm$0.12 \\
	\hline                                                                                                                                           
      \end{tabular}                                                                                                                               
\begin{flushleft}
$^a$ S\,Ori~J053811.0$-$023601 is at 28.0~arcsec from \object{S\,Ori~70}.
\end{flushleft}
\end{table*}                                                                                                                                      

Besides recovering the two faintest cluster member candidates\footnote{For the
latter, we were able to measure $H=18.8\pm0.2$~mag and $K_{\rm
s}=18.5\pm0.1$~mag, implying early L colours $J-H=0.8$~mag and $J-K_{\rm
s}=1.1$~mag.} \object{S\,Ori~J053932.4$-$025220} and
\object{S\,Ori~J054011.6$-$025135} from \citet{caballero2007} and the T-type
\object{S\,Ori~70}, we detect three new L- and T-type candidates and two
probable galaxies (see finding charts of Figs.~\ref{fcharts72}, \ref{fcharts73},
\ref{fcharts74glx1}, and \ref{fchartsglx2}), among many other objects rejected
because they do not meet our selection criteria. As shown in Fig.~\ref{IJ}, the
new candidates are about one magnitude fainter than S\,Ori~70. The photometric
information that we compiled from the images of different depths are listed in
Tables~\ref{LTcan}~and~\ref{gcan}, where the 5$\sigma$~flux upper limits
correspond to the magnitude limits of the images.

\begin{figure}[ht!] \resizebox{\hsize}{!}{\includegraphics{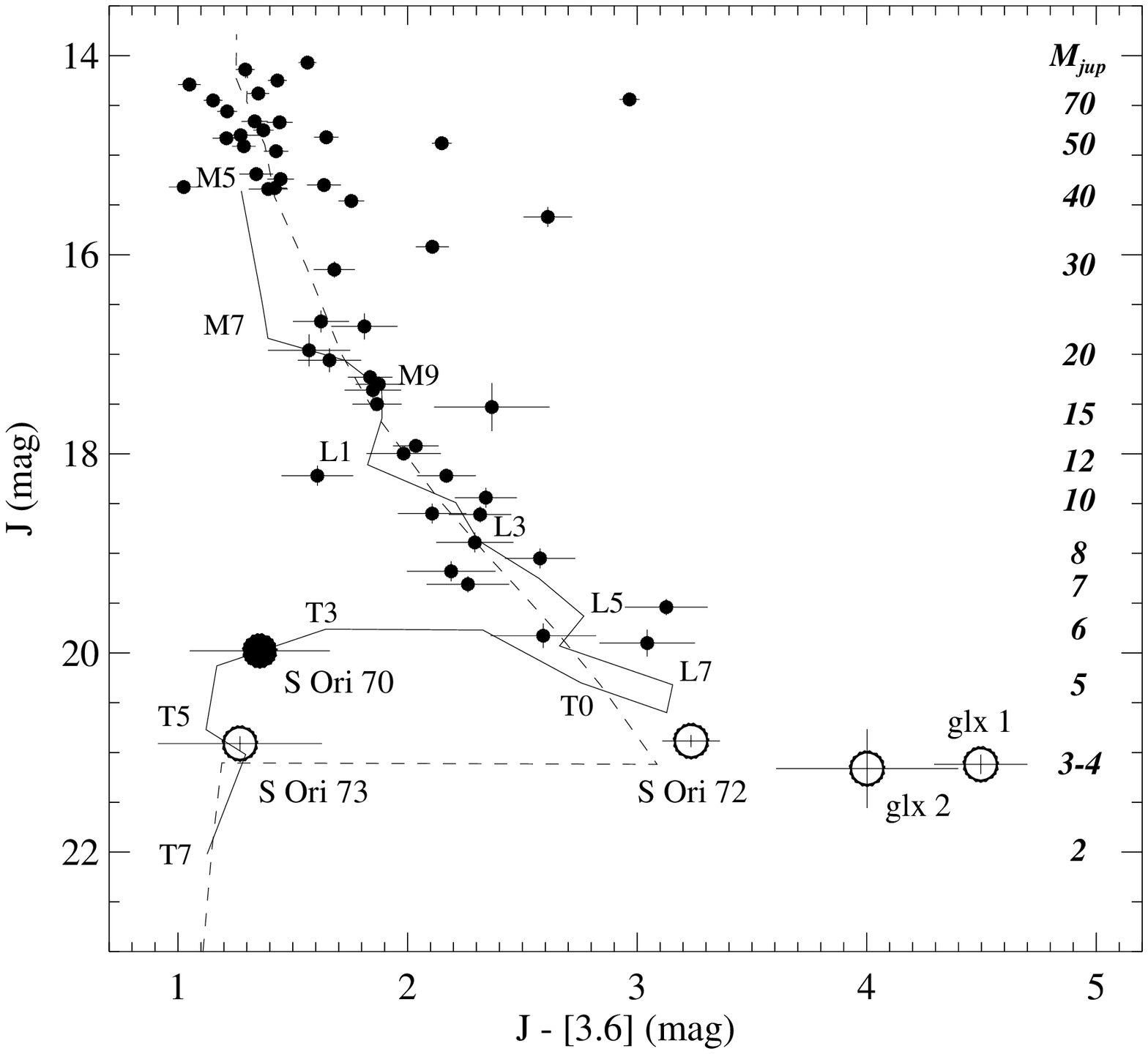}}
\caption{$J$ versus $J-[3.6]$ colour-magnitude diagram of $\sigma$~Orionis
cluster members and candidates. The filled circles correspond to objects from
\citet{caballero2007} and \citet{zapateroosorio2007,zapateroosorio2008}. The
large filled circle is S\,Ori~70. The large open circles are the new candidates.
S\,Ori~J053840.8$-$024022 and S\,Ori~J053811.0$-$023601 are labelled ``glx~1''
and ``glx~2'', respectively. The solid line represents the field dwarf sequence
shifted to match the brightness of the late-M-type $\sigma$~Orionis cluster
members. The dashed line represents the 3~Myr COND model isochrone (see text);
masses are indicated to the right in units of Jupiter masses.}  \label{3p6}
\end{figure}

\begin{figure}[ht!] \resizebox{\hsize}{!}{\includegraphics{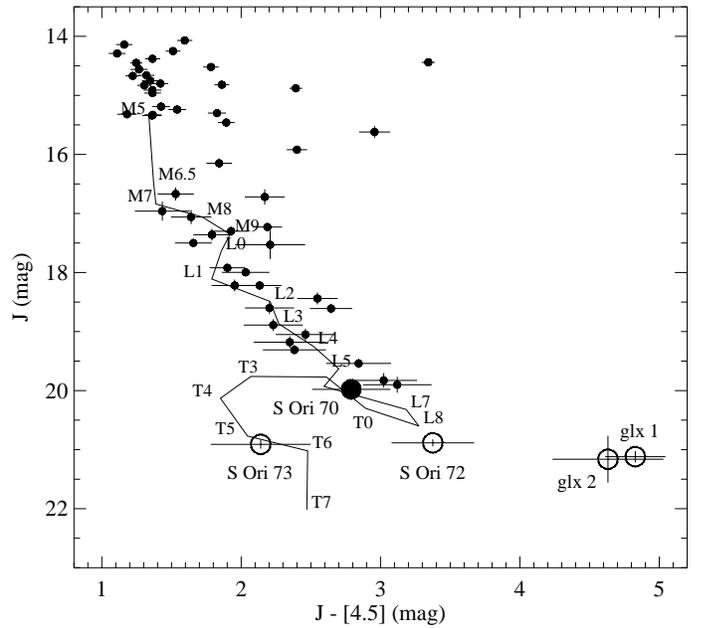}}
\caption{$J$ versus $J-[4.5]$ colour-magnitude diagram. Same as in
Fig.~\ref{3p6}.}  \label{4p5} \end{figure}

In the $J$ versus $J-[3.6]$ and $J-[4.5]$ colour-magnitude diagrams of
Figs.~\ref{3p6} and \ref{4p5}, we represent the candidates together with known
$\sigma$~Orionis cluster members and candidates
\citep{caballero2007,zapateroosorio2007,zapateroosorio2008}. The solid line
represents the spectrophotometric sequence of field mid-M- to late-T-type
dwarfs, shifted to match the brightness of the late-M-type cluster members
\citep{zapateroosorio2008}. For the field dwarfs, we use average absolute
$I$-band magnitudes, $I-J$, and $J-K_{\rm s}$ colours compiled by
\citet{caballero2008}\footnote{Note that the photometric values in Table 3
therein correspond to the spectral types M3V, M4V, M5V,... instead of M3-4V,
M4-5V, M5-6V,...The $I$ and $JK_{\rm s}$ magnitudes are in the Johnson-Cousins-
and 2MASS photometric systems, respectively.}, $J-H$ colours from
\citet{vrba2004}\footnote{The $J-H$ colour is transformed back from the CIT- to
the 2MASS photometric system using the same colour transformation of
\citet{carpenter2001} as used in \citet{vrba2004}.}, and mid-infrared magnitudes
from \citet{patten2006}. In Fig.~\ref{3p6}, the dashed line represents the 3~Myr
COND model isochrone at the cluster distance, adapted by converting predicted
effective temperature and luminosity into observables using relations for field
dwarfs \citep[procedure explained in][]{zapateroosorio2008}.

\begin{figure}[ht!] \resizebox{\hsize}{!}{\includegraphics{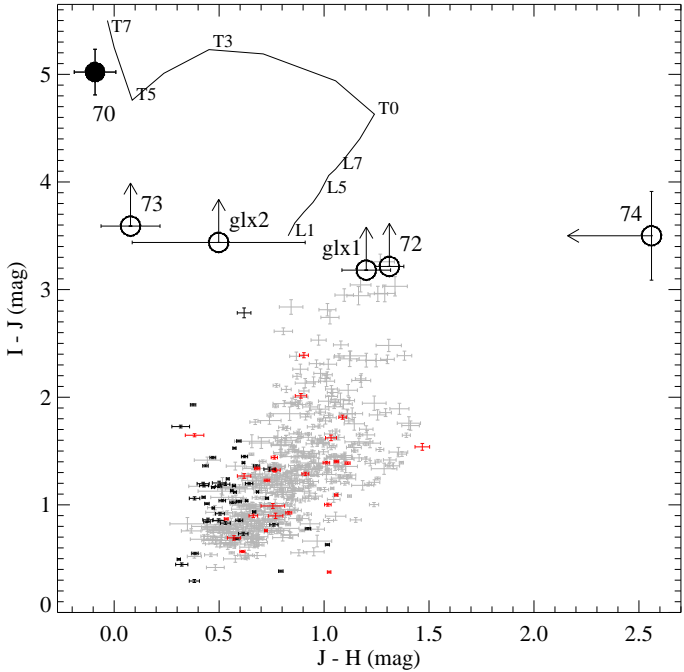}}
\caption{$I-J$ versus $J-H$ colour-colour diagram of stars (black crosses),
AGN (red crosses), and galaxies (grey crosses) from the GOODS-MUSIC catalogue
\citep{grazian2006}, in the magnitude range $19.5<J<21.5$~mag. The solid line
represents the field L1--T7-type dwarf sequence, the filled circle represents
S\,Ori~70, and the open circles represent the new candidates.
S\,Ori~J053840.8$-$024022 and S\,Ori~J053811.0$-$023601 are labelled ``glx~1''
and ``glx~2'', respectively.}  \label{ijhgoods} \end{figure}

\begin{figure}[ht!] \resizebox{\hsize}{!}{\includegraphics{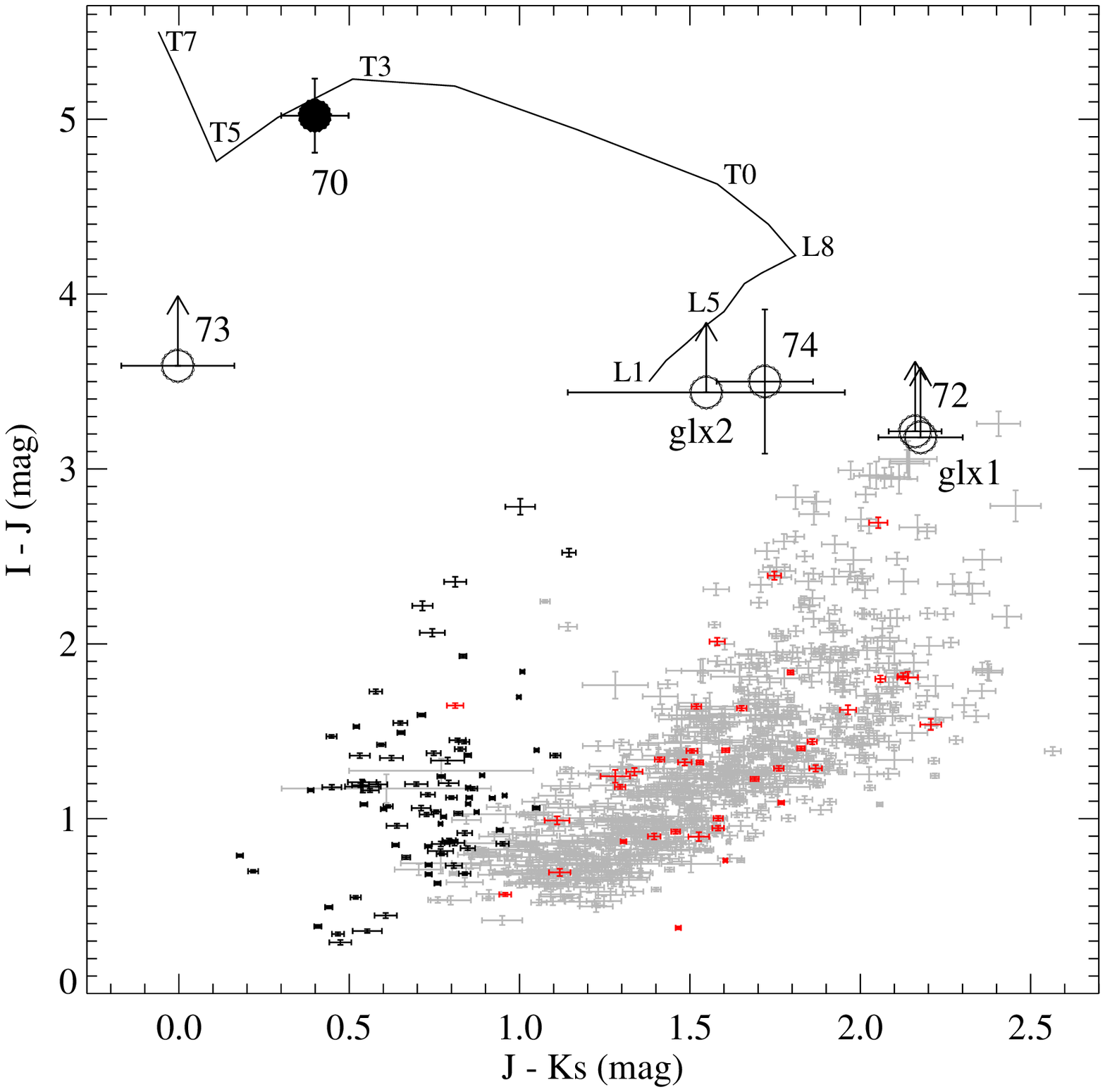}}
\caption{$I-J$ versus $J-K_{\rm s}$ colour-colour diagram. Same as in
Fig.~\ref{ijhgoods}.}  \label{ijksgoods} \end{figure}

\begin{figure*}[ht!] \sidecaption
\includegraphics[width=12cm]{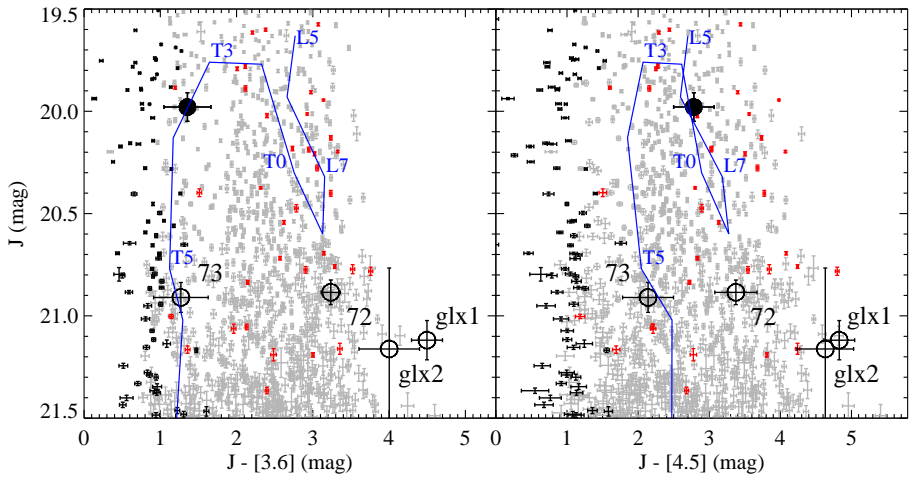} \caption{$J$ versus
$J-[3.6]$ and $J-[4.5]$ colour-magnitude diagrams. Same as in
Fig.~\ref{ijhgoods}. The blue solid line represents the field dwarf sequence
shifted as in Fig~\ref{3p6}.}  \label{j3p64p5goods} \end{figure*}

\begin{figure*}[ht!]
\sidecaption
\includegraphics[width=12cm]{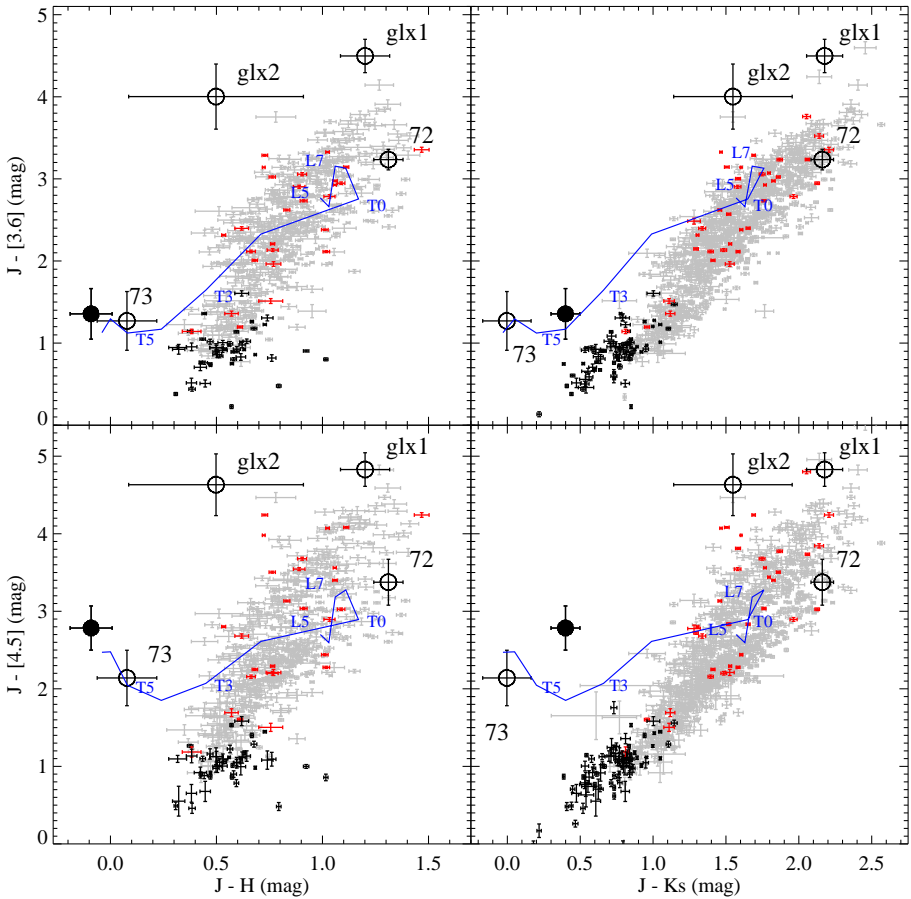}
\caption{Near- and mid-infrared colour-colour diagrams: $J-[3.6]$ and
$J-[4.5]$ versus $J-H$ (top- and bottom left), $J-[3.6]$ and $J-[4.5]$ versus $J-K_{\rm
s}$ (top- and bottom right). Same as in Fig.~\ref{ijhgoods}.} 
\label{nIRmidIRgoods} \end{figure*}

\begin{figure*}[ht!] \sidecaption
\includegraphics[width=12cm]{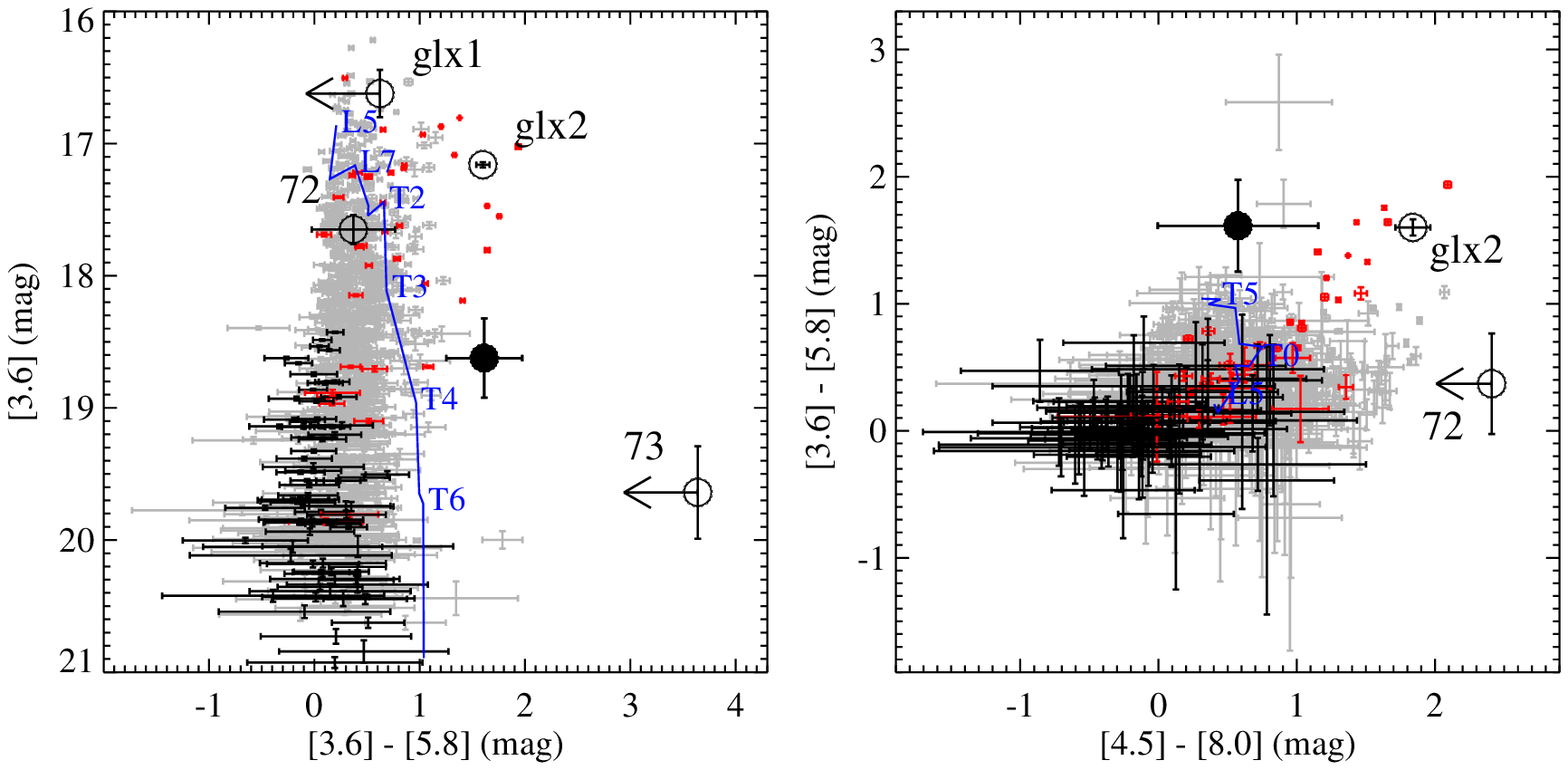}
\caption{Mid-infrared diagrams: $[3.6]$ versus $[3.6]-[5.8]$ (left),
$[3.6]-[5.8]$ versus $[4.5]-[8.0]$ (right). Same as in Fig.~\ref{ijhgoods}.} 
\label{midIRgoods} \end{figure*}

We also represent the candidates in $I-J$ versus $J-H$ and $J-K_{\rm s}$
colour-colour diagrams (Figs.~\ref{ijhgoods} and \ref{ijksgoods}) as well as in
various diagrams with mid-infrared filters (Figs.~\ref{j3p64p5goods},
\ref{nIRmidIRgoods}, and \ref{midIRgoods}), together with sources from the
GOODS-MUSIC catalogue \citep{grazian2006} that we use as a control field to
study the potential contamination by extragalactic sources in our survey. In all
the figures, the open circles represent the new candidates (labelled) and the
solid line represents part of the field LT-type dwarf sequence. The GOODS-MUSIC
survey is centred on $(\alpha, \delta)= (3~32~30, -27~48~30)$, approximately,
and covers an area of 143.2~arcmin$^{2}$ (except in the $H$-band, where the area
is 78~arcmin$^{2}$), i.e., less than a fifth of our search area. The limiting
magnitudes are $i=26.1$, $J=23.6$, $H=22.9$, $K_{\rm s}=21.9$, $[3.6]=21.2$,
$[4.5]=20.1$, $[5.8]=18.3$, and $[8.0]=17.6$~mag, when converted from the AB
system to the Vega system\footnote{The $i$- (or $F775W$) band photometry is from
$HST$/ACS and similar to that from the Sloan Digital Sky Survey (SDSS). A
transformation $I_{\rm Cousins}=i_{\rm SDSS}-0.3780*(i_{\rm SDSS}-z_{\rm
SDSS})-0.3974$ ($\sigma=0.0063$~mag) has been obtained for stars
(http://web.archive.org/web/20071014232413/http://www.sdss.org/
dr6/algorithms/sdssUBVRITransform.html); we assumed that this transformation is
also valid for galaxies. The VLT/ISAAC $JHK_{\rm s}$-band photometry was
converted to the Vega system using the transformations provided in the web page
http://web.archive.org/web/20070814141037/http://www.eso.org/
science/goods/releases/20050930/; it is found to be consistent within 0.05~mag
with the 2MASS point-source photometry. For the {\em Spitzer}/IRAC
$[3.6][4.5][5.8][8.0]$-band photometry, we used the transformations provided in
the web page http://web.ipac.caltech.edu/staff/gillian/cal.html.}. In the
magnitude range $19.5<J<21.5$~mag, the catalogue contains 882~galaxies,
37~active galactic nuclei (AGN), and 83~stars, and in the smaller $H$-band area,
505~galaxies, 27~AGN, and 49~stars. Stars and AGN are distinguished from
``normal'' galaxies mostly by morphological and photometric criteria, or else by
spectroscopic criteria. Stars are distinguished from AGN by spectroscopic
criteria. In the infrared colour-colour diagrams of Figs.~\ref{nIRmidIRgoods}
and \ref{midIRgoods}, the sequence of mid-L- to early-T~type field dwarfs
overlaps with the domain of galaxies and AGN; only T~type dwarfs tend to have
different colours. The colour-magnitude diagrams of Fig.~\ref{j3p64p5goods} also
indicate that, from $J=19.5$ to 21.5~mag, the colour ranges of galaxies and AGN
become broader and the number of galaxies increases (here by a factor of 1.4 in
a 0.5-magnitude interval). However the optical-infrared diagrams of
Figs.~\ref{ijhgoods} and \ref{ijksgoods} show that mid-L to mid-T~type dwarfs
are clearly redder in $I-J$ than the other sources. Thus, the $I-J$ colour is
essential to distinguishing these objects from galaxies and AGN, whereas the
infrared colours only help us to guess the spectral type.

\subsection{New L- and T-type candidates}

\begin{figure}[ht!] \resizebox{\hsize}{!}{\includegraphics{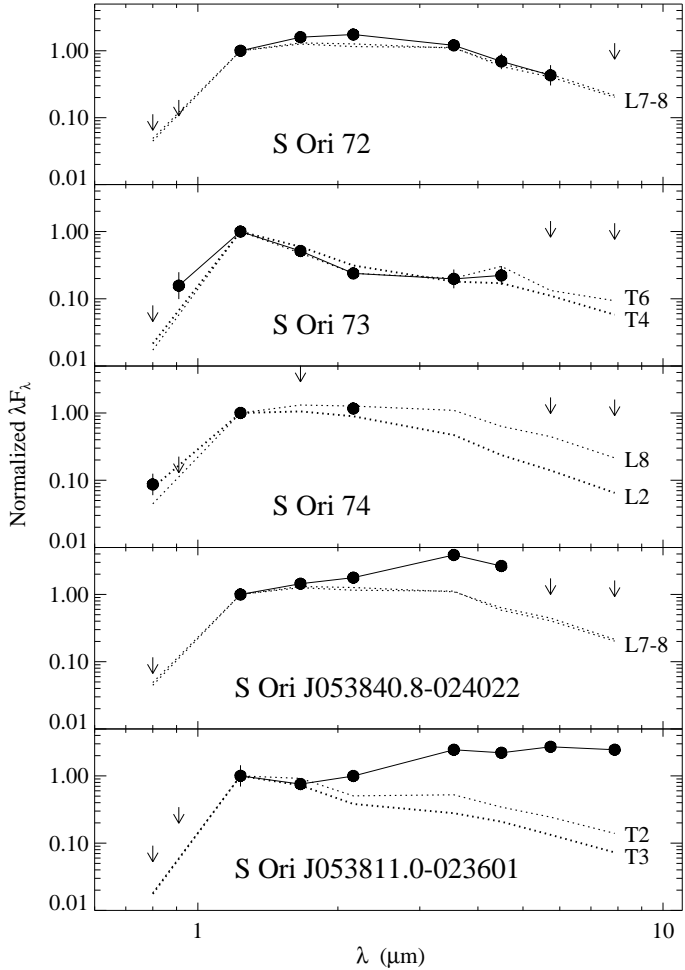}}
\caption{Spectral energy distributions of the new candidates compared
to average ones of field dwarfs (dotted lines). $IZJHK_{\rm
s}[3.6][4.5][5.6][8.0]$ from left to right.}  \label{seds} \end{figure}

{\it \object{S\,Ori~72}}, with $J-H\approx1.3$~mag and $J-K_{\rm
s}\approx2.2$~mag, could be a late~L-type object. It is clearly detected in the
ISAAC $J$-band image from December 2001 and even better in Omega2000 $HK_{\rm
s}$-band images (observations 2-h and 2-k, Table~\ref{obslog}), whereas it is
slightly blended but detected in the $[3.6][4.5][5.8]$-band images (its visual
neighbour is \object{2MASS~J05385930$-$0235282}, a field star located 3~arcsec
south-east, with $J\approx16.0$~mag and $I-J\approx0.9$~mag). S\,Ori~72 is
undetected in the WFC $I$-band image and the LRIS $IZ$-band images (1998iz3,
Table~\ref{IZlris}). The FWHMs in the $JHK_{\rm s}$-band images are
approximately equal to those of nearby faint point-like sources, of about 0.5,
1.2, and 0.8~arcsec, respectively. Among the new candidates presented in this
paper, S\,Ori~72 is the only one that is detected in the GCS-UKIDSS images, with
$K_{\rm UKIDSS}=18.60\pm0.28$~mag, in agreement with our measurement. In
Fig.~\ref{seds}, we show its spectral energy distribution, together with average
ones of field dwarfs of L7- and L8 spectral types (dotted lines). S\,Ori~72 is
relatively brighter in the $HK_{\rm s}$-bands. A preliminary measurement of
S\,Ori~72's proper motion using the ISAAC-Omega2000 images of 3.87~yr time
baseline and the method described in \citet{bihain2006} allows us to impose a $2
\sigma$ upper limit of 30 mas/yr. Although the estimate should be improved
before comparison with the $\mu\la5-10$~mas~yr$^{-1}$ amplitude of
$\sigma$~Orionis members \citep{caballero2007bri}, it indicates that S\,Ori~72
is not a high-proper motion object and unlikely to be a nearby ($\le$30~pc)
source. From the $J$ versus $J-[3.6]$ and $J-[4.5]$ colour-magnitude diagrams of
Figs.~\ref{3p6} and \ref{4p5}, S\,Ori~72 could be an L/T transition cluster
member candidate, but Figs.~\ref{ijhgoods}, \ref{ijksgoods},
\ref{nIRmidIRgoods}, and \ref{midIRgoods} imply that it could also be a galaxy
or an AGN.

With $J-H \sim J-K_{\rm s}\sim0$~mag, {\it \object{S\,Ori~73}} has near-infrared
colours of a mid T-type object. It is clearly detected in the ISAAC $J$-band
image, but appears very faint in the WFC $Z$-band image from November 2008, the
CFHTIR $HK'$-band images from February 2004 (20-h and 20-k,
Table~\ref{HKother}), and the public {\em Spitzer}/IRAC $[3.6]$- and
$[4.5]$-band images. It is undetected in the WFC $I$-band image, the LRIS
$IZ$-band images (2000i2, 1998iz2, Table~\ref{IZlris}), and in the Omega2000
$HK_{\rm s}$-band images (17-h and 17-k). Its FWHM in the $J$-band
image is approximately equal to that of nearby faint point-like sources, of
about 0.5~arcsec. In Fig.~\ref{seds}, we show its spectral energy distribution,
together with average ones of field dwarfs of T4 and T6 spectral types. The
position of S\,Ori~73 in the $J$ versus $J-[3.6]$ diagram of Fig.~\ref{3p6}
agrees with both adapted field- and model sequences, securing this source as a
good T-type- and cluster member candidate. The colour-colour diagrams of
Figs.~\ref{ijhgoods}, \ref{ijksgoods}, and \ref{nIRmidIRgoods} also indicate
that there are neither stars, nor AGN, nor galaxies in GOODS-MUSIC as red in
$I-J$ and blue in $J-H$ and $J-K_{\rm s}$ as this object.

With $I-J=3.5\pm0.4$~mag and $J-K_{\rm s}\approx1.7$~mag, {\it
\object{S\,Ori~74}} could be an L-type object. It is detected in the ISAAC
$J$-band image and in an Omega2000 $K_{\rm s}$-band image (18-k,
Table~\ref{HKother}). The object appears point-like in both images with a
typical local FWHM of 0.7 and 0.9~arcsec, respectively, whereas it appears faint
and blended with a stellar spike in the $[3.6][4.5][5.8]$-band images,
preventing us from determining its IRAC photometry. The candidate is barely
detected in the LRIS $I$-band images (2000i3 and 1998i3, Table~\ref{IZlris}) and
undetected in a stellar glare in the WFC $I$-band- and LRIS $Z$-band images
(1998z3). In Fig.~\ref{seds}, we show its spectral energy distribution, together
with average ones of field dwarfs of L2- and L8 spectral types. From
Fig.~\ref{ijksgoods}, it could also be a galaxy or an AGN. However, S\,Ori~74 is
located 11.8~arcsec north ($\sim$4250~AU; open square in Fig.\ref{map}) of the
bright K7.5-type cluster star \object{Mayrit~260182}, 4.3~arcmin south of the
cluster centre. The probability of chance alignment is only 5~\% for cluster
members at this angular distance from the $\sigma$ Orionis centre
\citep[][Fig.~1]{caballero2009asp}. Mayrit~260182, \object{Mayrit~270181}, and
\object{Mayrit~277181} also form a possible triple system \citep{caballero2006}.
\citet{caballero2006sori68} previously proposed that \object{S\,Ori~68} +
\object{SE 70} is a planet-brown dwarf system candidate, of greater mass ratio
and smaller separation. There are, however, known red galaxies at comparable
angular separations to $\sigma$~Orionis cluster members. For example, the Type~I
obscured quasi-stellar object \object{UCM~0536--0239} is located about
14.9~arcsec south of the T~Tauri star \object{Mayrit~97212}
\citep{caballero2008spec}. Observational follow-up is necessary to confirm
whether S\,Ori~74 is a cluster member and companion of Mayrit~260182.

\subsection{Probable galaxy candidates}

With $J-H\approx1.2$~mag and $J-K_{\rm s}\approx2.2$~mag, {\it
\object{S\,Ori~J053840.8$-$024022}} could be a late~L-type object or a galaxy.
It is detected with brighter magnitudes at longer wavelengths, from the
Omega2000 $JHK_{\rm s}$-band images (18j--k, Table~\ref{HKother}) to the
$[3.6][4.5]$-band images, but is then undetected in the $[5.8][8.0]$-band
images. It is undetected in the WFC and LRIS $I$-band images (2000i3,
Table~\ref{IZlris}). In Fig.~\ref{seds}, we show its spectral energy
distribution, together with average ones of field dwarfs of L7- and L8 spectral
types. In the ISAAC $J$-band image, it is found to be slightly extended and
fainter than in the lower-resolution Omega2000 $J$-band image. The FWHMs of the
object in the Omega2000 $JHK_{\rm s}$-band images are systematically larger by a
factor $\approx$1.4 than those of nearby faint point-like sources, suggesting
that it is a galaxy. S\,Ori~J053840.8$-$024022 could be an L/T transition
object, but from the $J$ versus $J-[3.6]$ and $J-[4.5]$ diagrams
(Figs.~\ref{3p6} and \ref{4p5}), it is redder than the expected sequence of the
cluster. Galaxies from the GOODS-MUSIC catalogue with these red colours appear
at magnitudes $J\ga20.5$~mag (Fig.~\ref{j3p64p5goods}).
Figure~\ref{nIRmidIRgoods} illustrates its infrared excess in the $K_{\rm s}$-,
$[3.6]$-, and $[4.5]$-bands relative to the field dwarf sequence, and also
suggests that this object is likely to be a galaxy.

{\it \object{S\,Ori~J053811.0$-$023601}}, with $J-H\sim0.5$~mag and $J-K_{\rm
s}\sim1.5$~mag, could be an early T-type object or a galaxy. It is detected with
brighter magnitudes at longer wavelengths, from the Omega2000 $JHK_{\rm s}$-band
images (18j--k, Table~\ref{HKother}) to the [3.6][4.5][5.6][8.0]-band images. It
is undetected in the WFC $I$-band image and the LRIS $IZ$-band images (2000i1
and 1998iz1, Table~\ref{IZlris}). In Fig.~\ref{seds}, we show its spectral
energy distribution, together with average ones of field dwarfs of T2 and T3
spectral types. We caution that the object appears relatively faint in the
$J$-band and that the $HK_{\rm s}$-band centroids are 0.8~arcsec south of the
$J$-band centroid, although the $JH$-band data were obtained on the same
observing night. The FWHMs in the $HK_{\rm s}$-band images are approximately
equal to those of nearby faint point-like sources, of about 1.2 and 0.9~arcsec,
respectively. S\,Ori~J053811.0$-$023601 appears as an L/T transition object, but
from the $J$ versus $J-[3.6]$ and $J-[4.5]$ diagrams (Figs.~\ref{3p6} and
\ref{4p5}) and as for \object{S\,Ori~J053840.8$-$024022}, it is redder than the
expected cluster sequence and could be a galaxy (Fig.~\ref{j3p64p5goods}).
S\,Ori~J053811.0$-$023601 is particularly bright in the $[5.8]$- and
$[8.0]$-bands and displays a colour $[3.6]-[8.0]=2.5\pm0.1$, redder than most
$\sigma$~Orionis low-mass member candidates
\citep{zapateroosorio2007,scholz2008,luhman2008}. Considering that $\ge$50\,\%
of the known $\sigma$~Orionis planetary-mass candidates exhibit excesses
longward of 5~$\mu$m \citep{zapateroosorio2007}, it appears to be a cluster
member. In Figs.~\ref{ijhgoods}, \ref{ijksgoods}, and \ref{nIRmidIRgoods}, its
optical and near-infrared colours differ from those of AGN and galaxies, but in
Fig.~\ref{midIRgoods}, its other colours are consistent with the AGN hypothesis
(see also Fig.~1 in \citealt{stern2005}, representing spectroscopically
identified stars, AGN, and galaxies). Hence, although we cannot exclude
completely this source beeing a peculiar cluster member with extreme infrared
excesses, our data seem to indicate that it is more probably an AGN.

\subsection{Cluster membership}\label{cmsd}

Because our search area is larger than that of the GOODS-MUSIC catalogue,
contamination by red galaxies and AGN is even more likely to explain some of our
candidates. \citet{caballero2008spec} present low-resolution optical
spectroscopy and spectral energy distributions between 0.55 and 24~$\mu$m of two
sources fainter than the star-brown-dwarf cluster boundary, which were
interpreted to be peculiar $\sigma$~Orionis members with very red colours
related to discs. They are instead two emission-line galaxies at moderate
redshift, one with an AGN and the other ongoing star formation. In the present
study, we assume that S\,Ori~J053840.8$-$024022 and S\,Ori~J053811.0$-$023601
are galaxy- or AGN contaminants and that the other objects are Galactic
candidates awaiting confirmation by higher resolution imaging, proper motion, or
spectroscopy.

In our search, we must also account for contamination by field dwarfs.
\citet{caballero2008} provide predictions of the number of L5--T0, T0--T5, and
T5--T8 field dwarf contaminants per square degree towards the $\sigma$~Orionis
region, in one-magnitude $I$-band intervals and from $I=21.0$ to 29.0~mag. We
convert the bright and faint boundaries of the range $J=19.7-21.1$~mag (as a
prolongation of the search range of \citealt{caballero2007}) into the $I$-band
magnitudes corresponding to the earliest and latest spectral types of each of
the three contaminant groups. We then sum the predicted numbers of contaminants
accounting for the $I$-band ranges and scale the sums to the search area that is
complete to $J\ge21.1$~mag ($\sim$790~arcmin$^{2}$). We obtain about three
L5--T8-type field dwarfs, which all contribute the most to the light close to
$J=21.1$~mag. This predicted value remains mostly indicative, because the
initial mass function and scale heights of late~L- and T-type dwarfs are still
uncertain. Interestingly, \citet{caballero2008} assume a rising mass function in
the planetary mass regime and predict spatial densities of T0--8~dwarfs that are
a factor of two higher than those derived from observations
\citep{metchev2008,lodieu2009farT}.

S\,Ori~73 and S\,Ori~70 are located\footnote{S\,Ori~73 is the only candidate
found in one of the deeper multi-band search areas, summing up to
$\sim$470~arcmin$^2$, see Sect.~\ref{tsearch}.} at 11.9 and 8.7~arcmin from
$\sigma$~Ori~AB, respectively. S\,Ori~72 and S\,Ori~74 are closer, at 3.6 and
4.1~arcmin, respectively. Interestingly, the location of these faintest,
presumably least massive candidates contrasts with that of the eleven
13--6~$M_{\rm Jup}$ free-floating planetary-mass candidates from
\citet{caballero2007}, further out at 26--13~arcmin in the survey area (see
Fig.~\ref{map}). \citet{caballero2008dis} find an apparent deficit of low mass
objects ($M<0.16$~M$_\odot$) towards the $\sigma$~Orionis cluster centre. If the
cluster membership census and the individual masses are confirmed, this
configuration could be explained by several mechanisms, including e.g., a
possible photo-erosion by the central OB stars \citep{hester1996,whitworth2004}
in the deep gravity well. Complementary studies of the dense cluster core
\citep{bouy2009} and other cluster regions could thus help us to understand the
formation of low-mass planetary-mass objects.

\subsection{Mass spectrum}

We consider the luminosity and mass functions for the ISAAC- and additional
areas, where the search is complete down to $J\ge21.1$~mag
($\sim$790~arcmin$^{2}$).

\begin{figure}[ht!] \resizebox{\hsize}{!}{\includegraphics{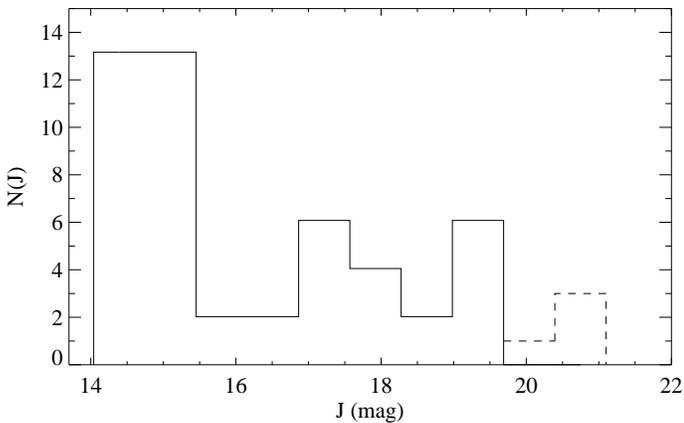}}
\caption{$J$-band luminosity function with the LT-type candidates at
$J>19.7$~mag (dashed line) and the brighter cluster member candidates from
\citet{caballero2007} scaled to the search area (solid line).}
\label{lf} \end{figure}

In Fig.~\ref{lf}, we show the $J$-band luminosity function. The magnitude bins
in the range $J=19.7$--21.1~mag correspond to the three new LT-type candidates
and S\,Ori~70 (dashed line). The magnitude bins in the range $J=14.1$--19.7~mag correspond to
the cluster member candidates from \citet{caballero2007}, i.e., in the ISAAC
area; they are scaled by the area factor $(790)/780=1.0128$. The magnitude
bins have equal widths of about 0.7~mag.

\begin{figure}[ht!] \resizebox{\hsize}{!}{\includegraphics{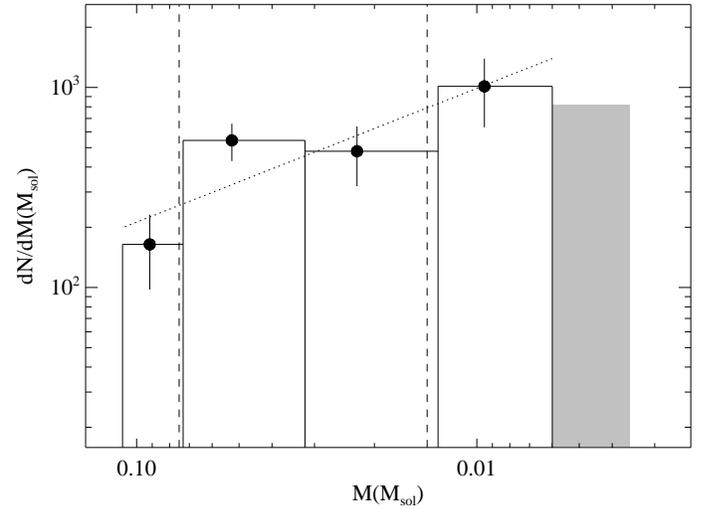}}
\caption{Mass spectrum with contamination-corrected data. The dotted segment
represents the linear fit to the data points from \citet{caballero2007} in the
mass range 0.11--0.006~$M_{\odot}$, which are previously scaled to the total
area. The shaded region is our estimate of 0--2 cluster members in the mass
range 0.006--0.004~$M_{\odot}$. From left to right, the vertical dashed lines
represent the hydrogen and deuterium burning mass limits, respectively.} 
\label{mf} \end{figure}

We estimate the masses of our new cluster member candidates by comparing with
the theoretical bolometric luminosities from the Lyon group
\citep[e.g.,][]{baraffe2003}, using exactly the same method as in
\citet{caballero2007}. If cluster members, S\,Ori~72--74 would each have an
estimated theoretical mass of 4$_{-2}^{+3}$~$M_{\rm Jup}$, accounting for age,
distance, and photometric uncertainties. This rounded up result does not change
significantly by using a cluster distance of 400~pc \citep{mayne2008} or 440~pc
\citep{sherry2008} instead of 360~pc \citep{brown1994}. The effective
temperature corresponding to that mass would be of $\sim$1400~K. In
Fig.~\ref{mf}, we display the mass spectrum ($\Delta N/\Delta M$). The filled
circles represent the contamination-corrected data points from
\citet{caballero2007} scaled to the search area of $\sim$790~arcmin$^{2}$. The
last bin (shaded region) corresponds to the result from the present study for
the magnitude range $J=19.7$--21.1~mag. Subtracting the three possible
contaminants (see Sect.~\ref{cmsd}) from the four LT-type candidates and
accounting for the Poissonian error, we estimate 0--2 cluster members with a
mass of 0.006--0.004~M$_{\odot}$.

Previous studies of the substellar population in the $\sigma$~Orionis cluster
find that the mass spectrum increases toward lower masses. \citet{bejar2001}
show that it can be represented by a potential law ($\Delta N/\Delta M \propto
M^{-\alpha}$) with an $\alpha$ index of 0.8 in the mass range
0.11--0.013~M$_{\odot}$. \citet{gonzalezgarcia2006} and \citet{caballero2007}
extend this mass spectrum to 0.006~M$_{\odot}$ and find a slightly lower index
$\alpha=0.6$. For the substellar mass range of 0.073--0.006~$M_{\odot}$,
\citet{caballero2007} obtain an even lower $\alpha$ index of 0.4. An
extrapolation of the mass spectrum with an index $\alpha=0.4-0.8$ predicts 3--7
objects in the mass range 0.006--0.004~M$_{\odot}$. From our survey, the most
likely number of  cluster members in this mass interval is in the range 0--2.
This could be an indication of a turnover in the substellar mass spectrum.
However, given the low statistics and the possibility that the number of
contaminants could be overestimated, such a change in the slope of the mass
spectrum should be considered with caution. If real, the turnover could be
related to an opacity mass limit, turbulence effects, or a different
mass-luminosity relation (if less massive objects were fainter than predicted).
Wider and deeper searches would be very valuable in constraining the mass
spectrum more reliably at these and lower masses.

\section{Conclusions}

The mass function in young open clusters can provide clues about the formation
mechanism of free-floating planetary-mass objects. We therefore decided to
explore the substellar mass function for $M<6$~$M_{\rm Jup}$ in the $\sim$3~Myr
old \object{$\sigma$~Orionis} open cluster. We extended to $J=19.5-21.5$~mag the
$\sim$780~arcmin$^{2}$ INT/WFC-VLT/ISAAC $IJ$-band search of
\citet{caballero2007}. $J$-band sources (ISAAC and CAHA 3.5~m/Omega2000) were
cross-matched with $I$- (WFC and Keck/LRIS) and $HK$-band sources (Omega2000,
NTT/SofI, WHT/LIRIS, and CFHT/CFHTIR). We selected sources redder than a
boundary at $I-J>3.1-3.5$ or without an $I$-band detection or fainter than
$I=24$~mag. These sources were then checked visually in all available images,
including $Z$-band images from LRIS and WFC, and archival mid-infrared images
from {\em Spitzer}/IRAC.

We recover \object{S\,Ori~70} and the two faintest cluster member candidates
from \citet{caballero2007}, and we find five red $I-J$ sources, with
$J\sim21$~mag, located within 12~arcmin of the cluster centre. The near- and
mid-infrared colours indicate that one of the sources, \object{S\,Ori~73}, is
probably of T~spectral type. If confirmed as a cluster member, it would be the
least massive free-floating T~type object detected in $\sigma$~Orionis, with
4$_{-2}^{+3}$~$M_{\rm Jup}$. The four other sources appear to be L/T transition
objects, but two are likely to be galaxies because of their strong mid-infrared
excesses, similar to those of galaxies at $J\ga20.5$~mag. \object{S\,Ori~72} and
S\,Ori~73 are relatively close to the expected cluster sequence in the $J$
versus $J-[3.6]$ and $J-[4.5]$ colour-magnitude diagrams. \object{S\,Ori~74} is
located 11.8~arcsec ($\sim$4250~AU) away from the solar-type cluster star
\object{Mayrit~260182}. From the effective search area of $\sim$790~arcmin$^{2}$
complete to $J=21.1$~mag, we estimate there to be, after contaminant correction,
between zero and two cluster members in the mass interval 6--4~M$_{\rm Jup}$.
The low number of candidates in this mass bin may be indicative of a turnover in
the substellar mass function. Wider and deeper optical-to-infrared surveys are
required to confirm whether this is the case, by constraining the mass function
more tightly at lower masses.

\begin{acknowledgements}

We thank the referee Kevin Luhman. We thank Claire Halliday (A\&A language
editor) and Terry Mahoney (IAC, Spain) for revising the English of the
manuscript. We acknowledge Project No. 03065/PI/05 from the Fundaci{\'o}n
S{\'e}neca. Partially funded by the Spanish MEC under the Consolider-Ingenio
2010 Programme grant CSD2006-00070 (First Science with the GTC,
http://www.iac.es/consolider-ingenio-gtc/). Based on observations made with ESO
Telescopes at the La Silla or Paranal Observatories under programmes ID
068.C-0553(A) and 078.C-0402(A). Based on observations obtained at the
Canada-France-Hawaii Telescope (CFHT), which is operated by the National
Research Council of Canada, the Institut National des Sciences de l'Univers of
the Centre National de la Recherche Scientifique of France, and the University
of Hawaii. Based on observations collected at the German-Spanish Astronomical
centre, Calar Alto, jointly operated by the Max-Planck-Institut f{\" u}r
Astronomie Heidelberg and the Instituto de Astrof\'{\i}sica de Andaluc\'{\i}a
(CSIC). We thank Calar Alto Observatory for allocation of director's
discretionary time to this programme. Based on observations made with the Isaac
Newton Telescope (INT) and the William Herschel Telescope (WHT) operated on the
island of La Palma by the Isaac Newton Group in the Spanish Observatorio del
Roque de los Muchachos of the Instituto de Astrof\'{\i}sica de Canarias. This
research has been supported by the Spanish Ministry of Science and Innovation
(MICINN) under the grant AYA2007-67458. Some of the data presented herein were
obtained at the W.M. Keck Observatory, which is operated as a scientific
partnership among the California Institute of Technology, the University of
California, and the National Aeronautics and Space Administration. The
Observatory was made possible by the generous financial support of the W.M. Keck
Foundation. The authors wish to recognise and acknowledge the very significant
cultural role and reverence that the summit of Mauna Kea has always had within
the indigenous Hawaiian community. We are most fortunate to have the opportunity
to conduct observations from this mountain. This work is based in part on
observations made with the Spitzer Space Telescope, which is operated by the Jet
Propulsion Laboratory, California Institute of Technology under a contract with
NASA. {\tt IRAF} is distributed by the National Optical Astronomy Observatories,
which are operated by the Association of Universities for Research in Astronomy,
Inc., under cooperative agreement with the National Science Foundation. This
publication makes use of data products from the Two Micron All Sky Survey, which
is a joint project of the University of Massachusetts and the Infrared
Processing and Analysis centre/California Institute of Technology, funded by the
National Aeronautics and Space Administration and the National Science
Foundation. This research has made use of SAOImage DS9, developed by Smithsonian
Astrophysical Observatory. This research has made use of the SIMBAD database,
operated at CDS, Strasbourg, France

\end{acknowledgements}

\bibliographystyle{aa}
\bibliography{12210.bbl}

\appendix
\section{Representation of individual fields in the $IZJHK$-bands}

\begin{figure*}[ht!] \resizebox{\hsize}{!}{\includegraphics{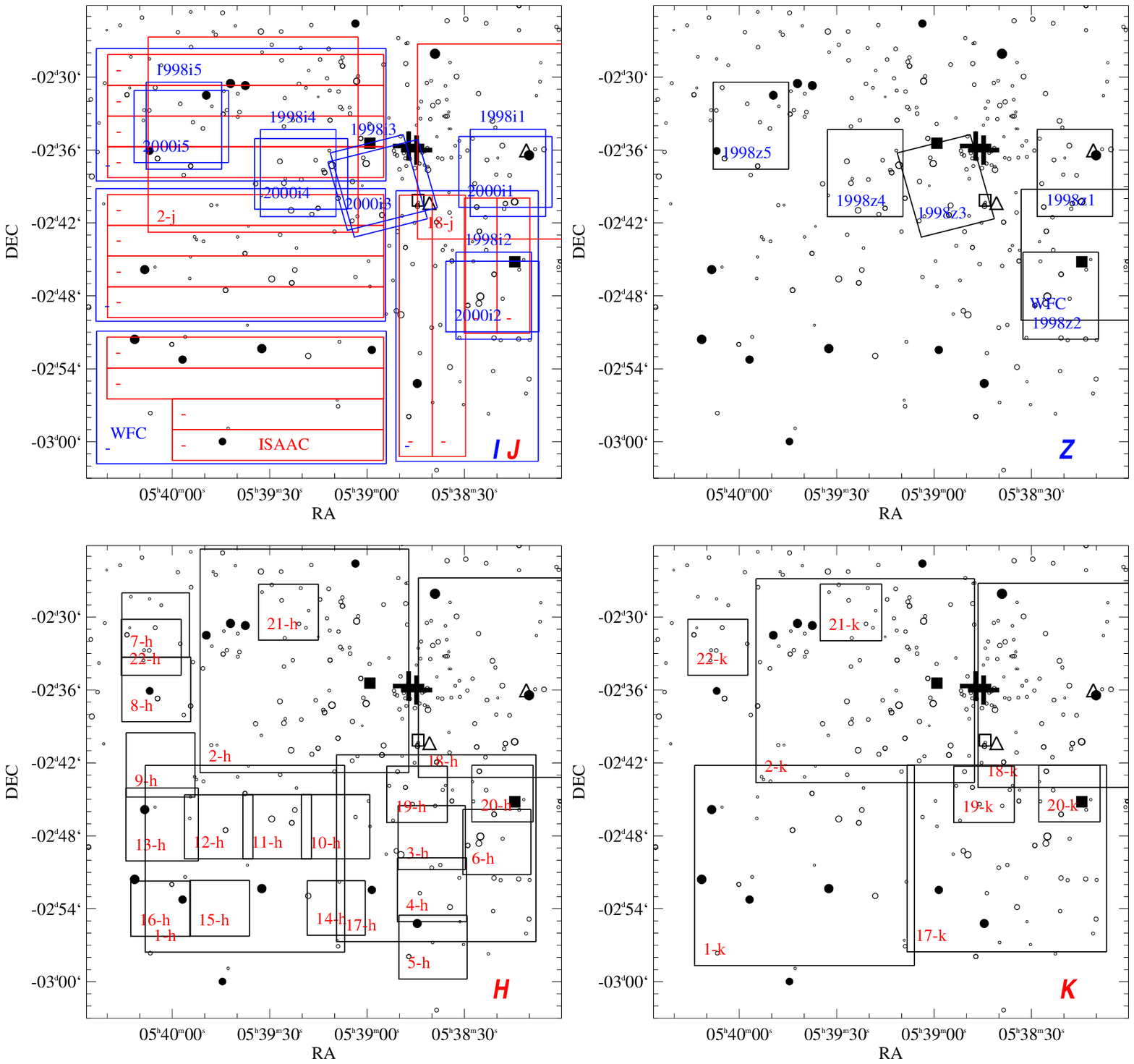}}
\caption{Individual fields used in the search: $IJ$-bands ({\it top
left}), $Z$-band ({\it top right}), $H$-band ({\it bottom left}), and $K$-band
({\it bottom right}). Symbols are defined as in Fig.~\ref{map}.}  \label{mapsi}
\end{figure*}

\section{Finding charts}

  \begin{figure}[ht!]                                                                                                                               
\resizebox{\hsize}{!}{\includegraphics{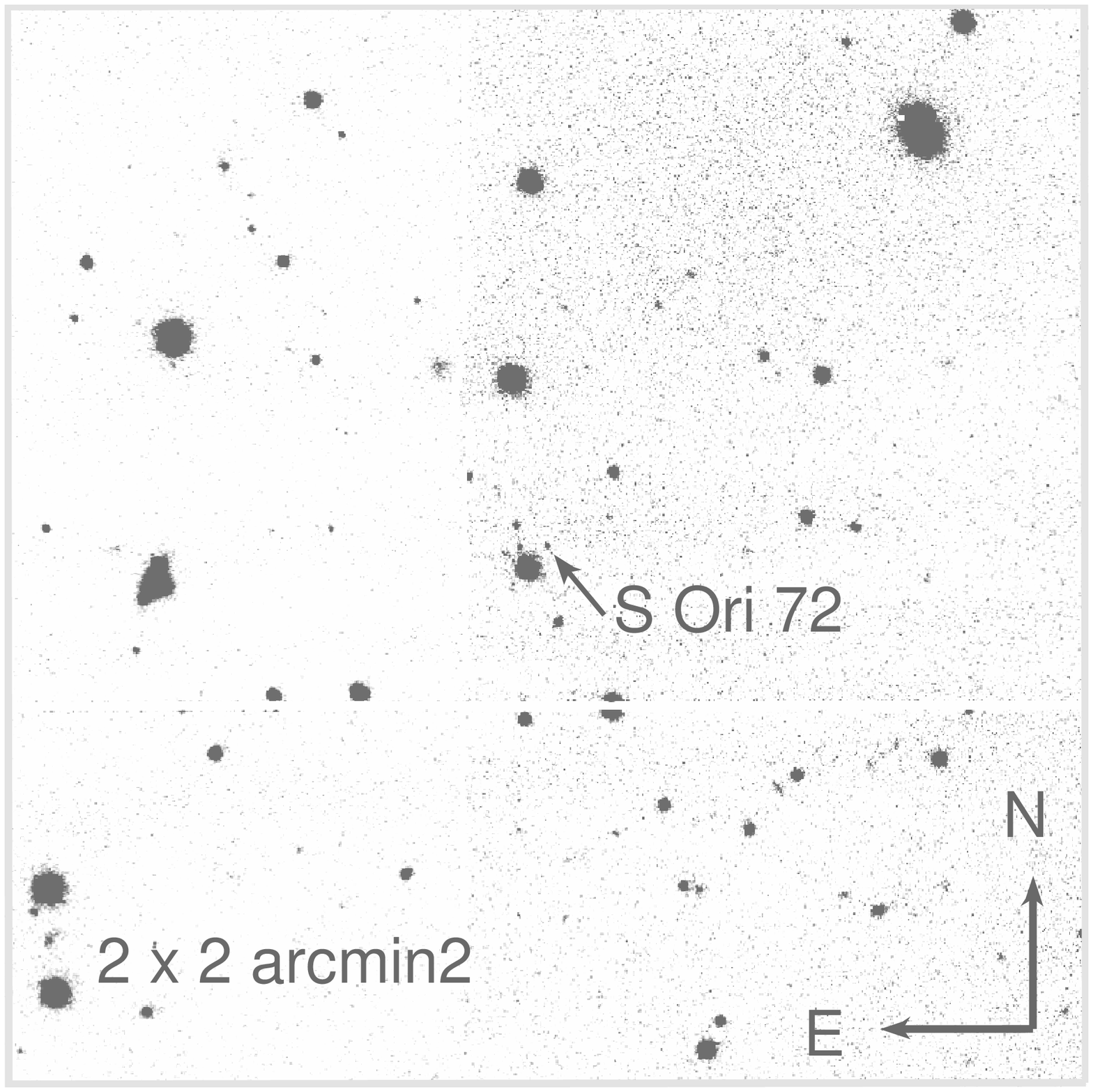}}								      
\caption{ISAAC $J$-band image of S\,Ori~72.}                                                                                              
\label{fcharts72}
\end{figure}                                                                                                                  

  \begin{figure}[ht!]                                                                                                                               
\resizebox{\hsize}{!}{\includegraphics{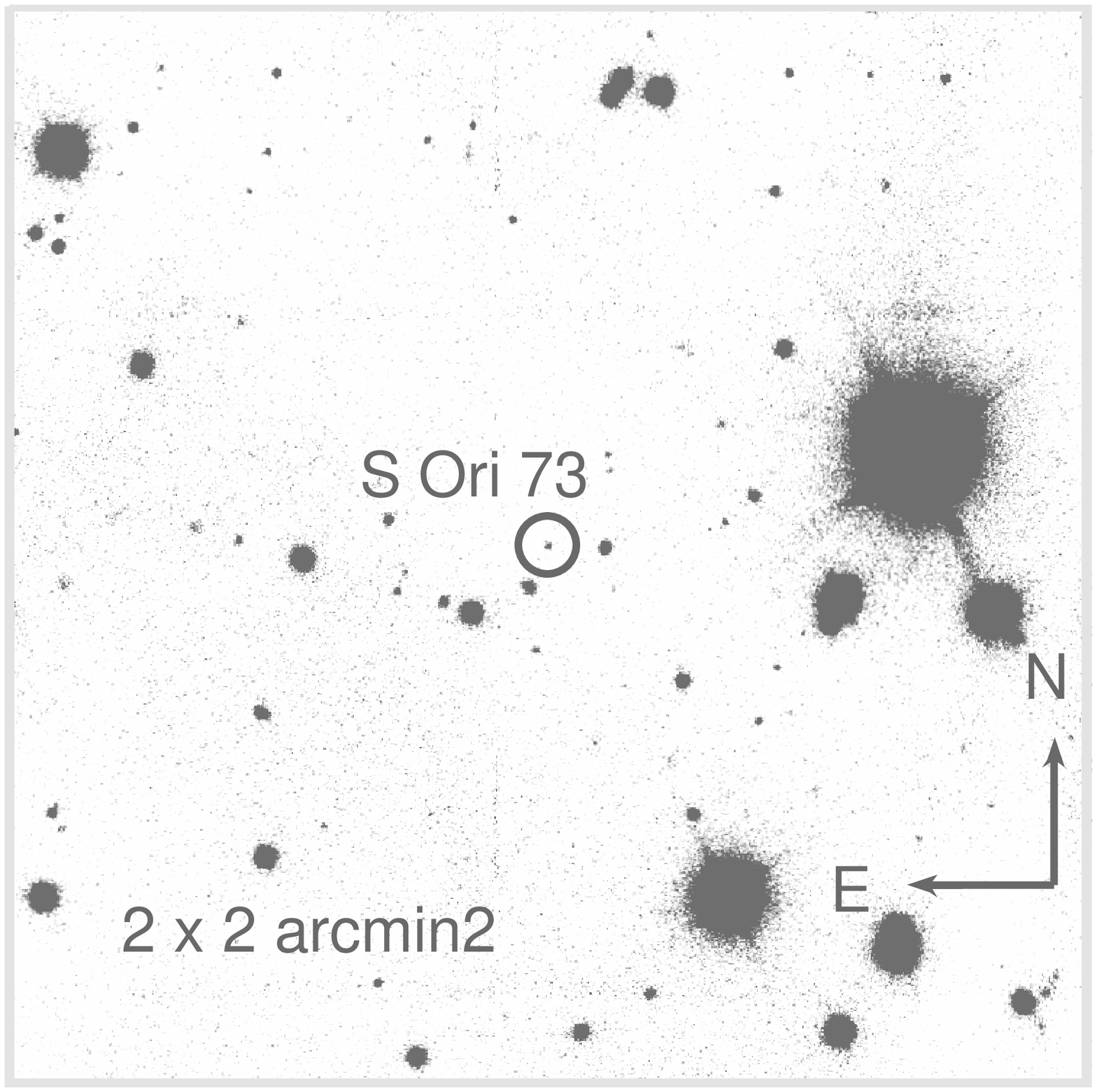}}								      
\caption{ISAAC $J$-band image of S\,Ori~73.}                                                                                              
\label{fcharts73}
\end{figure}                                                                                                                  

  \begin{figure}[ht!]                                                                                                                               
\resizebox{\hsize}{!}{\includegraphics{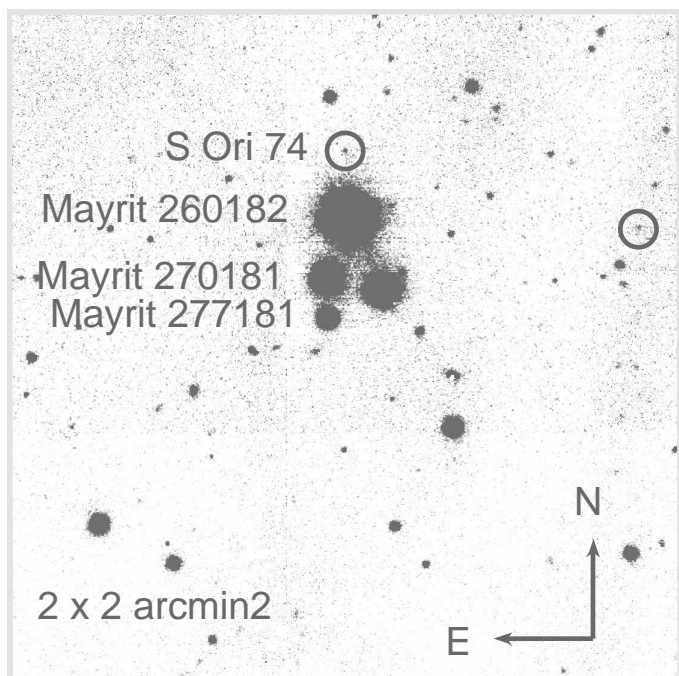}}								      
\caption{ISAAC $J$-band image of S\,Ori~74 and S\,Ori~J053840.8$-$024022 (unlabelled circle).}                                                                                              
\label{fcharts74glx1} \end{figure}                                                                                                                  

  \begin{figure}[ht!]                                                                                                                               
\resizebox{\hsize}{!}{\includegraphics{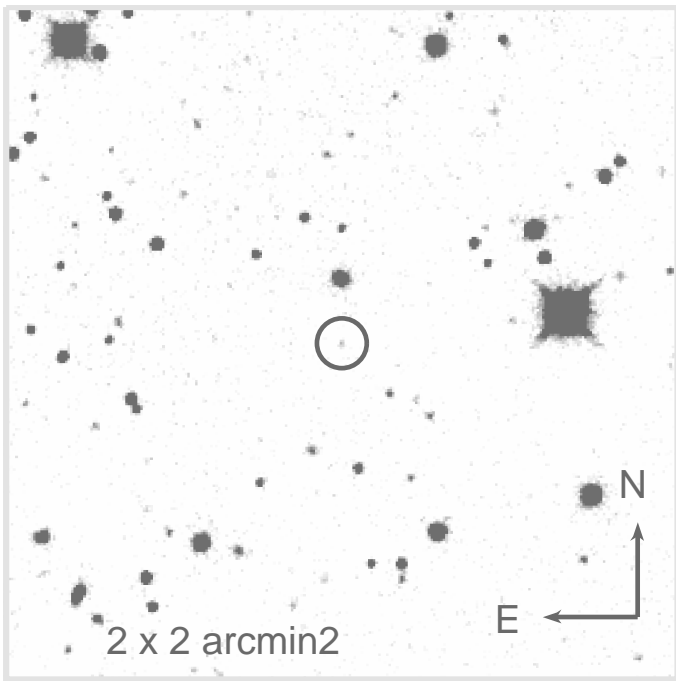}}								      
\caption{Omega2000 $H$-band image of S\,Ori~J053811.0$-$023601.}                                                                                              
\label{fchartsglx2} \end{figure}

\end{document}